\documentstyle{l-aa}     % LaTeX A&A  Standard Fonts, NORMAL PRINTOUT
\nonstopmode
\newcommand{\aua}{{\it Astron. Astrophys.} }

\newcommand{\auar}{{\it Astron. Astrophys. Rev.} }

\newcommand{\apj}{{\it Astrophys. J.} }
\newcommand{\apjl}{{\it Astrophys. J. Letters} }
\newcommand{\apjs}{{\it Astrophys. J. Suppl. Series} }
\newcommand{\aj}{{\it Astron. J.} }
\newcommand{\mn}{{\it Monthly Notices Roy. Astron. Soc.} }

\newcommand{\ltsim}{\lower 0.6ex\hbox{$\stackrel{\textstyle <}{\sim}$}}
\newcommand{\gtsim}{\lower 0.6ex\hbox{$\stackrel{\textstyle >}{\sim}$}}

\newcommand{\kms}  {$\mbox{km\,s}^{-1}$}
\newcommand{\beqy} {\begin{eqnarray}}
\newcommand{\beq}  {\begin{equation}}
\newcommand{\eeqy} {\end{eqnarray}}
\newcommand{\eeq}  {\end{equation}}
\newcommand{\cR}   {corotation}
\newcommand{\CR}   {Corotation}
\newcommand{\Cs}   {Contopoulos}
\newcommand{\PA}   {\mbox{P.A.}}
\font\caps=cmcsc10			   %caps & small caps (IAU etc)
\def\maybebreak{\vskip 0pt plus .1\hsize \penalty -250
                \vskip 0pt plus -.1\hsize}
\def\maybesmallbreak{\vskip 0pt plus .07\hsize \penalty -250
                \vskip 0pt plus -.07\hsize}
\newcount\chapnum
\newcount\secnum
\newcount\subsecnum
\newcount\eqnum
\def\thechapnum{\the\chapnum}
\def\chap#1\par{\maybebreak \bigskip\bigskip\bigskip\bigskip
                \global\advance\chapnum by 1
                \global\eqnum=0 \global\secnum=0 \global\subsecnum=0
                \noindent{\bf\thechapnum.\ #1}
                \medskip\noindent}
\def\sec#1\par{\ifvmode\maybesmallbreak\bigskip\bigskip\fi
               \global\advance\secnum by 1
               \global\subsecnum=0
               \noindent{\caps\thechapnum.\the\secnum\ #1}
               \nobreak\smallskip\noindent}
\def\subsec#1\par{\ifvmode\maybesmallbreak\bigskip\fi
                  \global\advance\subsecnum by 1
                 \noindent{\bf\thechapnum.\the\secnum.\the\subsecnum
                              \quad#1}}

\def\spose#1{\hbox to 0pt{#1\hss}}
\def\lta{\mathrel{\spose{\lower 3pt\hbox{$\mathchar"218$}}
     \raise 2.0pt\hbox{$\mathchar"13C$}}}
\def\gta{\mathrel{\spose{\lower 3pt\hbox{$\mathchar"218$}}
     \raise 2.0pt\hbox{$\mathchar"13E$}}}
\def\ssim{\!\sim\!}

\def\tkms{{\rm\,km\,s^{-1}}}

\def\kpc{{\rm\,kpc}}

\def\gyr{{\rm\,Gyr}}

\def\as{{\arcsec}}
\begin{document}

\thesaurus{11
             (11.09.1 M 94;
              11.11.1;
              11.16.1;
              11.19.2)
          }
%\thesaurus{11.09.1 M 94;11.11.1;11.16.1;11.19.2}

\title{The central bar in M\,94}

% \subtitle{}

\author{C.\ M\"ollenhoff
\inst{1}
\thanks{Visiting astronomer of the German-Spanish
Astronomical Center, Calar Alto, operated by the Max-Planck-Institut f\"ur
Astronomie, Heidelberg jointly with the Spanish National Commission for
Astronomy.}
\and M. Matthias\inst{1,2}
\and O. E. Gerhard\inst{1,2} }

\offprints{C.\ M\"ollenhoff}

\institute{Landessternwarte, K\"onigstuhl 12, D-69117 Heidelberg, Germany\\
e-mail: cmoellen@mail.lsw.uni-heidelberg.de \and
Astronomisches Institut, Universit\"at Basel, Venusstr. 7,
CH-4102 Binningen, Switzerland \\
e-mail: matthias@astro.unibas.ch, gerhard@astro.unibas.ch}

\date{Received Aug. 1994; accepted }

\maketitle

%--------------------------------------------------------------------------
\begin{abstract}
%______________________________________ Do not leave a blank line here!
Visual, NIR, and $H_{\alpha}$ surface photometry of the oval disk
galaxy M\,94 (NGC\,4736) was performed to study the distribution of
mass in old stars and the distribution of the warm emission line gas.
The radial profile of the inner stellar disk is steeper than
exponential and displays twisted and deformed isophotes. Dust is not
responsible for these deviations. The investigation of a number of
morphological models showed that M\,94 has a weak central stellar bar
of 0.7 kpc major axis length which comprises about 14\% of the total
light within 20\as.

By means of longslit spectroscopy the kinematics of gas and stars in
this region were investigated.  The stellar kinematics reveal the
existence of a small spheroidal bulge with $v/\sigma\approx 0.8$.  The
stellar velocity field in this region is approximately axisymmetric,
showing that the effects of the bar on the kinematics of the hot
component are relatively small. The warm gas in the bar region shows
global and local deviations from the stellar kinematics, but outside
the central bar fits into the general $HI$ velocity field.

Model calculations of closed orbits for gas in the potential of the
bar perturbation and axisymmetric disk and bulge predict large
non-circular motions for cold gas in equilibrium flow. These do not
fit the central $H_{\alpha}$ kinematics; rather it appears that
the $H_{\alpha}$ gas must have large random motions and is not
in a steady state, and that hydrodynamical effects due to a recent
starburst play a role in the central region.

\end{abstract}

%--------------------------------------------------------------------------
%--------------------------------------------------------------------------
\section{Introduction}
\label{intro}

M\,94 = NGC\,4736 is an Sab spiral galaxy at a distance of
$\approx6.6$ Mpc. It is well studied in optical morphology, optical
emission gas kinematics (van der Kruit 1974, 1976, Buta 1984, 1988)
and $HI$ distribution and kinematics (Bosma et al 1977, Mulder \& van
Driel 1993). There exist several good overviews about these results
and other characteristics of M\,94 in the literature (e.g. Buta 1984,
Mulder \& van Driel 1993). Since this paper deals mainly with new
optical and NIR results (morphology, stellar kinematics, gas
kinematics) we present here only a brief summary of the most important
optical features of this galaxy. Basic parameters of M\,94 and the
most numerical results from this study are presented in Table 1.

Bosma et al (1977) distinguished the following five annular zones
with progressivly lower surface brightness in M\,94:

\begin{enumerate}
\item A bright central region (bulge), $R < 15$\arcsec.
\item A zone of inner spiral structure $R \approx 15$ to $50$\arcsec.
      This zone is bounded by a bright inner ring.
\item A zone of outer spiral structure, $R \approx 50$ to $\approx 200$\arcsec,
      this region probably forms an oval disk.
\item A zone of low surface brightness ({\it gap}).
\item A faint outer ring of $R \approx 330$\arcsec.
\end{enumerate}

The nuclear region is extremely bright; according to Keel and Weedman
(1978) M\,94 is one of the galaxies with the highest optical surface
brightness.  The inner ring of radius $\approx 40${\arcsec}to
$50$\arcsec, shows $HII$ regions and young blue objects, characteristic
for strong star formation.  The interpretation of this ring is the subject
of numerous papers. In a detailed study of the kinematics van der Kruit
(1974, 1976) attempted to fit a variety of kinematical models to the
ring. He concluded that the velocity field is best interpreted in
terms of a uniform radial expansion from the nucleus. However, using
Fabry-Perot data Buta (1984, 1988) concluded that the ring is better
interpreted as the result of a secular evolution in this galaxy due to
a Lindblad resonance.

The bright inner ring was also observed in CO by Gerin et al (1991). They
followed the formation of the outer and the inner ring in a model where
the main disk of M\,94 is a massive oval with $b/a=0.8$. That the disk of
M\,94 is not axially symmetric had already been seen by Bosma et al (1977).

The zone inside the inner ring displays a strong isophote twist.
This could be the consequence of a triaxial bulge (Beckman et al 1991)
or a central secondary bar (Kormendy 1993). Such a 'bar within a bar'
would rotate much faster than the large outer oval disk (Shlosman et
al 1989, Friedli and Martinet 1993).  A third interpretation was given
by Shaw et al (1993), who proposed that this structure is the result
of a combined stellar- and gas-dynamical mechanism. In the potential
of the rotating oval disk a gas ring formas which
perturbs the central stellar
distribution.  This pertubation is misaligned with respect to the
outer oval disk but has the same angular speed; thus it is not an
independent bar.

The aim of this paper is to study the distribution and kinematics
of stars and gas in the central region of NGC\,4736,
and to compare these with model calculations.  In
Sect.~\ref{obser} we describe the observations: (1) Surface photometry
in the visual region ($V,I,H_{\alpha}$) and in the NIR ($J,K$), in
order to study the distribution the old stellar component as well as
that of the dust and warm, line-emitting gas. (2) Longslit
spectroscopy for the investigation of the kinematics of stars and gas
in this region. The data reduction and the results are described In
Sect.~\ref{result}. In Sect.~\ref{modmorph} we present morphological
models for M\,94. In Sect.~\ref{modkin} orbit families in a corresponding
model potential are calculated and are used to model the kinematics
of stars and gas. Their predictictions will be compared with the
observed kinematics.  In Section~\ref{discuss} we discuss our results,
and the conclusions of this paper are given in Section~\ref{concl} .

%--------------------------------------------------------------------------
%--------------------------------------------------------------------------
\section{Observations}
\label{obser}

%--------------------------------------------------------------------------
\subsection{CCD surface photometry}
\label{obsccd}

The CCD imaging observations were performed in February 1990, in
January 1992, and in July 1993 at the German-Spanish Astronomical
Center on Calar Alto, Spain.  The Calar Alto 1.23m telescope was used
in three different setup modi for CCD imaging: (1) The
Landessternwarte focal reducer at the Cassegrain focus with a GEC
P8603/A chip ($576 \times 386$ pixels, 1 pix of $22 \mu \cor
1.56$\arcsec) and $V, I$ filters; (2) the Calar Alto Cassegrain focus
CCD camera with a coated GEC chip ($576 \times 386$ pixels, 1 pix of
$22 \mu \cor 0.46$\arcsec) and $I$ filter; (3) the Cassegrain focus
CCD camera with a Tektronix chip ($512 \times 512$ pixels, 1 pix of
$27 \mu \cor 0.56$\arcsec) and $H_{\alpha}$ filter. The exposure times
were 1 minute ($I$ filter) and 3 minutes ($V$) with the focal reducer,
15 minutes ($I$) and 45 minutes ($H_{\alpha}$ filter) with the
Cassegrain camera. Photometric standard stars were exposed several
times during the night for the flux calibration.  Flat field exposures
were taken during evening and morning dusk; the telescope was slightly
displaced between each exposure, to allow the elimination of stellar
images in the flat field frames by an appropriate filtering procedure.

%--------------------------------------------------------------------------
\subsection{NIR surface photometry}
\label{obsnir}

The NIR imaging observations were performed in January 1994 with the
2.2m telescope of the German-Spanish Astronomical Center on Calar
Alto, Spain.  $J,K$ images were obtained with the MPIA MAGIC camera
(Herbst et al 1993).  MAGIC has a 256x256 NICMOS3 HgCdTe detector
array and was used in an imaging mode of 0.66\arcsec/pix,
corresponding to a field of 169{\arcsec}x169\arcsec. 3 exposures of 20
sec were added in the camera electronics, using a 'sample up the ramp'
readout mechanism. Then the telescope was moved by a few arcsec for
the next exposure, and another few arcsec for a third exposure, in
order to get rid of the bad pixels. In total six such triples, i.e. 18
images of 1 min exposure time were obtained in each filter.  Between
the galaxy exposures the telescope was moved approx. half a degree
offset for 3 blank sky images of the same expore times, also shifted
slightly from each other to remove background stars.  Photometric
standard stars were exposed for the flux calibration, the sky
exposures serve as flat fields.

%--------------------------------------------------------------------------
\subsection{Kinematics of the stars}
\label{obskins}

The spectroscopic observations were done during February 1990 with the
2.2m telescope of the German-Spanish Astronomical Center on Calar
Alto, Spain.  Longslit spectrograms were obtained using the standard
Boller {\&} Chivens spectrograph in 4 slit orientations, $P.A. =
0^{\circ}, 45^{\circ}, 90^{\circ}, 135^{\circ}$ respectively. The
detector was an RCA CCD with 1024 x 640 pixels of $15 \mu$ size. The
slit width was $250 \mu$ or 2.9\arcsec.

For the stellar kinematics we used a grid with 1200 lines/mm in the
wavelength range 4500 -- 5400 {\AA} (resolution 0.9 \AA/pix). 3 pixels
(or 2.7\arcsec) along the slit were binned during readout. The
exposure times were $2 \times 1 h$ for each slit position.

Several stellar templates of spectral type between G8 III and K5 III were
exposed for the subsequent Fourier-Cross-Correlation. The internal He-Ar lamp
was used for wavelength calibration. A large number of domeflats with different
exposure times were exposed for the correction of the nonlinear columns of the
RCA chip at low light levels. Skyflats were obtained to correct for the
profile along the slit.

%--------------------------------------------------------------------------
\subsection{Kinematics of the gas}
\label{obsking}

The spectrograms for the kinematics of the emission gas of the $HII$
regions were obtained during the same observing run with the same slit
width and orientations as above. The grid for the gas kinematics had
1200 lines/mm in the wavelength range 6200 -- 7100 {\AA}, and the
exposure times here were 30 min for each spectrogram. The full spatial
resolution of 0.9 \arcsec/pix along the slit was used.

During a second spectroscopic run in January 1994
gas spectrograms in further slit orientations $P.A. = 28^{\circ}, 73^{\circ},
118^{\circ}, 163^{\circ}$ respectively,  were obtained with the same
spectrograph and grating at the same telescope. However, this time a
Tektronix chip 1024 x 1024 of $24\mu$ pixelsize, slit width
$370\mu$ or 4.3\arcsec, and spatial resolution of 1.47\arcsec/pix was used.
The above given slit orientations correspond to the position angle of the
central bar in M\,94 (Sect.~\ref{resccd}).

%--------------------------------------------------------------------------
%--------------------------------------------------------------------------
\section{Data reduction and results}
\label{result}

%--------------------------------------------------------------------------
\subsection{CCD surface photometry}
\label{resccd}

The CCD images were reduced using the standard procedure with bias subtraction
and flatfield division. An intensity-dependant median filter was applied
which smoothed the images only in the outer zones of low S/N ratio
but not in the central region of the galaxy.

\begin{figure}
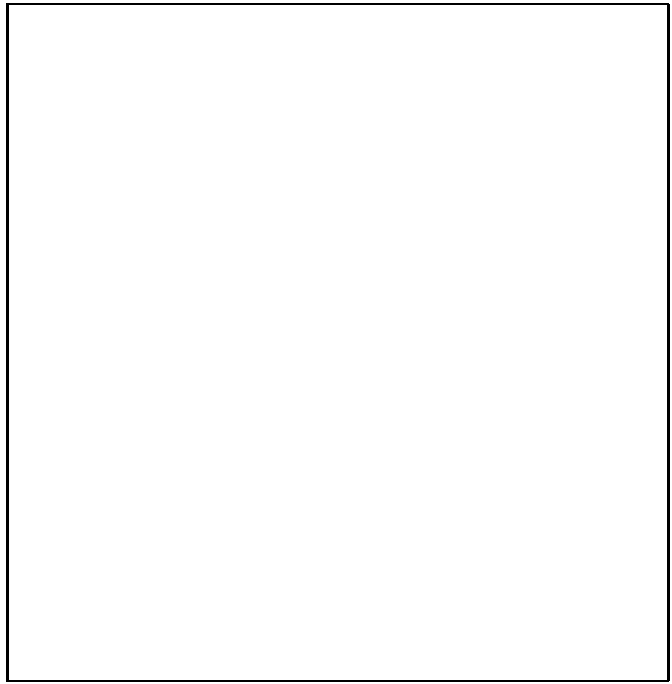

\picplace{9.0cm}
\caption[]{
Central region of M\,94 (NGC\,4736), $I$ filter, $78 \times 78$ arcsec,
north is up, east to the left. A logarithmic saw-tooth lookup table
was used to produce this image, which shows isocontours as dark lines
and the shallow gradients in between as different gray shadings. The
cusped deformations of the isophotes due to the central bar at
$P.A. = 28^{\circ}$ are easily visible.}
\label{figi}
\end{figure}

Figure~\ref{figi} shows the central $2 \times 2$ arcmin of the $I$
image of M\,94. The slightly irregular outer structure is the well
known oval ring of $HII$ regions, not very pronounced in this deep red
image. While the major axes of the outer isophotes at $r\approx
40${\arcsec} have a position angle ($P.A.$) $\approx 120^{\circ}$,
they are nearly circular at $r\approx 20${\arcsec}. Further inside the
position angle changes to $\approx 28^{\circ}$, and the isophotes
display a distinctly {\it cusped} structure.  Beckman et al (1991)
interpreted the isophote twist as a signature for a triaxial bulge.
However, the occurrence of cusped isophotes makes it more
plausible that a bar-like structure of $\sim30\arcsec$ = 1.0 kpc
length is superposed onto the inner disk of M 94. This bar was already
mentioned in Kormendy (1993).

A third interpretation was given by Shaw et al (1993), who proposed
that this structure is the result of a combined stellar- and
gas-dynamical mechanism: The oval disk leads to an accumulation of gas
in a non-axisymmetric circumnuclear ring which becomes phase-shifted with
respect to the original stellar oval disk. This ring has sufficient mass to
induce a triaxial perturbation in the central stellar component.
This perturbation
is misaligned with respect to the outer oval disk but has the same
angular speed; thus it is not an independent bar.

How does the isophote twist model of Shaw et al (1993) compare with
our observations? Shaw et al give only rather general statements
about the shape of the isophotes to be expected; it is difficult to
make a comparison with the observed morphological details.
They interprete the region of isophote shift as a perturbation
due to the (phase shifted) gas ring; thus it should extend out to the
radius of the ring.  However, the most important point is that the bar ends at
a finite radius {\it inside} the ring (cusped isophotes). Therefore we
consider the interpretation of Shaw et al as less plausible.

In Sect.~\ref{modmorph} we will present models of the central photometry
and discuss the nature of the barred component in more detail. We will give
arguments that it is indeed a bar. In the following we will therefore call it a
'bar' for (semantic) simplicity.

The I image was calibrated using the standard stars (M\,67) and comparing
with the aperture photometry from de Vaucouleurs \& Longo (1988).

%--------------------------------------------------------------------------
\subsection{NIR surface photometry}
\label{resnir}

The data reduction of the $J,K$ images proceeded as follows:
To each triple of sky exposures a vertical median filter was
applied, thus removing the stars. These sky frames were obtained
for each galaxy and each filter, they were subtracted from the 18 galaxy
images respectively. Since the dark current of the MAGIC camera
is very low (Herbst et al 1993) it was not corrected for.
The internal accuracy between {\it all} sky exposures of one night in
each filter is better than 0.2\%, with larger deviation from night to
night. Therefore the flat field images were constructed for every
night separately, using all sky images in each filter of that
night. After flat field division of the 18 galaxy frames the relative
offsets of the galaxy centers and of stars (if present) were
measured. Then the images were added together pixelwise, applying
their individual offsets. An internal comparison allowed us
to reject the bad pixel values. The $J,K$ images look very similar
to the $I$ image (Fig.~\ref{figi}), showing the same behaviour of the
isophotes and the bar feature. This is another support for the statement
that the bar is not a feature due to dust absorbtion.

%--------------------------------------------------------------------------
\subsection{Isophote fits}
\label{isoph}

\begin{figure*}
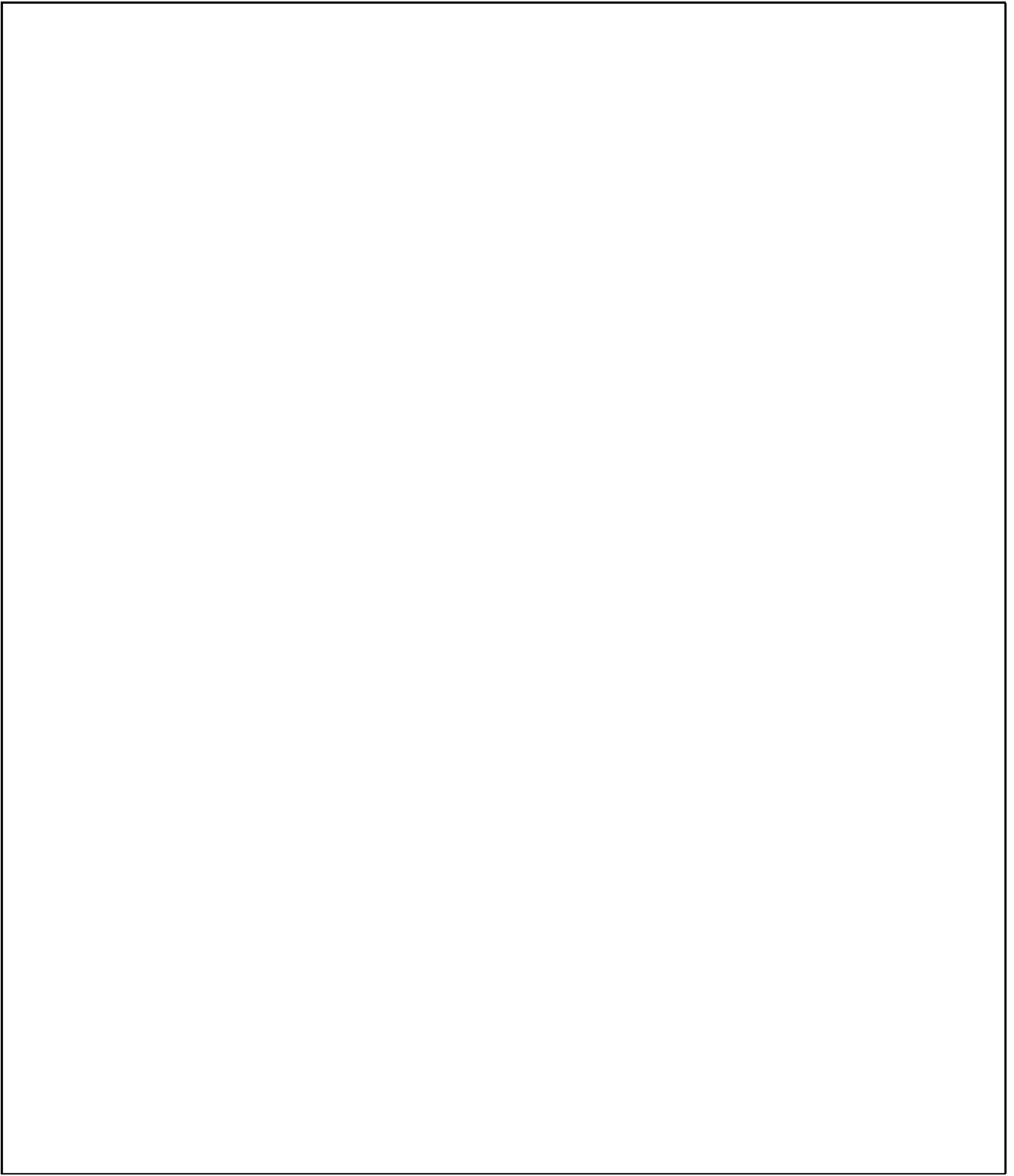

\picplace{21.0cm}
\caption[]{
Results of the isophote fits to the $K$ image (left panel) and the
focal reducer $I$ image (right panel). The plots show surface
brightness, ellipticity, position angle, and the $a_4$-Fourier
coefficient over radius. The central bar causes a first maximum
of the ellipticity at $\approx10${\arcsec} radius with a position angle
$P.A. = 28^{\circ}$ (left panel, $K$). For $r > 45${\arcsec} (outer
disk) the ellipticity increases slowly from 0.2 to 0.4, the
$P.A.$ decreases from $120^{\circ}$ to $90^{\circ}$ (right panel, $I$).
The structural changes in the disk at $\sim45${\as} and
$\sim120${\as} can be seen in all plots.}
\label{figell}
\end{figure*}

After masking of disturbing field stars the isophotes of the $I,J,$
and $K$ images were fitted by ellipses (Bender \& M\"ollenhoff,
1987). The results for the $K$ image and the $I$ focal reducer image
are shown in Fig.~\ref{figell} in the left and right panels
respectively. The upper plots show the surface brightness in the inner
part (left panel, $K$, $r<90$\arcsec), and out to the largest radius
we could obtain (right panel, $I$, $r<450$\arcsec).  As already stated
by Boroson (1981) and Kormendy (1982), M\,94 shows a steep
non-exponential profile in the inner disk region ($r<200${\arcsec}).
An exponential profile can be fitted only in the very outer
disk. However, this region of steeper profile is still a disk in the
kinematical sense: as will be seen later from the stellar velocity
dispersion data, the real bulge of M\,94 extends only to $\approx
20${\arcsec} radius.  The steep disk profile is not totally smooth, it
shows two 'shelves' at $r\approx45${\arcsec} and
$r\approx120${\arcsec} which, according to Kor\-men\-dy (1982) are
characteristics of oval disks. These 'shelves' can be detected in all
our visual and NIR surface brightness profiles.

The next plots in Fig.~\ref{figell} show ellipticities $\varepsilon =
1-b/a$ and position angles of the isophote fits to the $K$ and $I$
images. The curves are very similar in the different colors which
means that dust does not play a role in M\,94
(cf. Sect.~\ref{col}). Due to the bar structure the ellipticity
profile has a first maximum of $\varepsilon = 1-b/a = 0.22$ at
$\approx10${\arcsec} radius (Fig.~\ref{figell}, left panel), the
position angle is $P.A. = 28^{\circ}$ there.  In the transition zone
between central bar and outer disk at $r\approx 20${\arcsec} radius
the isophotes are nearly round ($\varepsilon \simeq 0$).  Then
$\varepsilon$ increases again, accompanied by a strong variation of
$P.A.$ between 20{\as} and 30{\as}. With increasing radius, the
ellipticity increases generally from 0.2 to 0.25 at $r \approx
120$\arcsec, and to $\sim0.4$ at large radii.  The general slope of
$1-b/a$ is only interupted by two local maxima in the ellipticity at
50{\as} and 125{\as}, followed by a slight local decrease of
$\varepsilon$. These maxima correspond to the two shelves of the
surface brightness at 45{\as} and 120{\as}. The main point here is,
that $\varepsilon \approx 0.18$ between $r = 50$ and 100{\as} and
$\varepsilon \approx 0.23$ between $r = 120$ and 220{\as}.  As
Kormendy (1982) mentioned, these nested elliptical regions of
different $\varepsilon$ indicate, that at most one of these zones can
be axial symmetric.  Since warps can be excluded at this high surface
brightness (Fig.~\ref{figell}), part of the disk must be oval.  If we
assume the zone between $r = 50$ and 100{\as} as axial symmetric, we
get an incliation angle of $i = 35^{\circ}$. Then the outer zone
between $r = 120$ and 220{\as} must be intrinsically elliptic with an
axial ratio of $b/a \approx 0.9$.

The bottom plots of Fig.~\ref{figell} show the $a_4$-Fourier
coefficent which quantifies symmetric deviations of the real isophotes
from ellipses (Bender \& M\"ollenhoff 1987). A positive $a_4$ value
marks cusped isophote deviations; they are especially strong at $r =
10${\arcsec} (left panel, $K$).  $a_4$ is large and positive in the
limited radius between $r = 5${\arcsec} and 15{\arcsec}; this suggests
that the bar ends at $r \approx 15${\arcsec}, well inside the star
formation ring at $r \approx 45$\arcsec. Then $a_4$ shows an
oscillating structure with increasing radius; at 45{\as} and 110{\as}
we observe a change from slightly negative $a_4$ values (i.e. boxy
deviations) to positive values. This finding is another confimation
of the structural changes in the disk of M\,94 at $r\approx
45${\arcsec} and $r \approx 120${\arcsec} (right panel, $I$).

%--------------------------------------------------------------------------
\subsection{Fit of the surface brightness profile}
\label{sbprof}

To study the light distribution in the inner part of M\,94 we used the
$I$ Cassegrain focus image, since with 0.46{\arcsec}/pix it has a
better pixel-resolution than the $J,K$ images with 0.66{\arcsec}/pix.
However, the $I$ image is still smoothed by atmospherical seeing,
which was determined from stellar images to a value of 1.76{\arcsec},

In first test fits of the surface brightness a de Vaucouleurs profile
was found too steep while a Hernquist (1990) profile already gave a
fairly good fit. For a better modelling we proceeded in a more general
way, making use of the $\gamma$--profiles described by Dehnen
(1993). They are characterized by a central luminosity density slope
$\sim r^{-\gamma}$ and a length scale $\alpha$. They include profiles
between a steep Jaffe type ($\gamma = 2$), de Vaucouleurs type
($\gamma = 1.5$), Hernquist type ($\gamma = 1$), to a flat core type
profile ($\gamma = 0.5$).

\begin{figure}
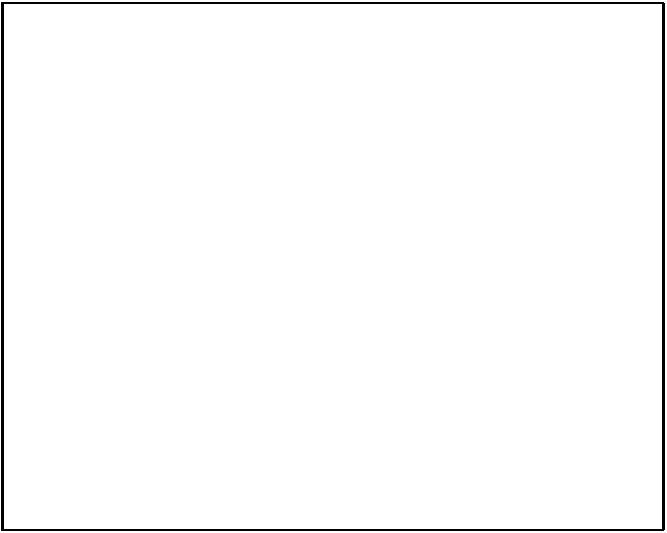

\picplace{7.0cm}
\caption[]{
Seeing-convolved $\gamma$--profiles with $\alpha = 15${\arcsec} for different
$\gamma$--values compared to the observed $I$--profile.
The curve for $\gamma = 0.9$ gives the best fit for $r \le 40 $\arcsec.}
\label{figgam}
\end{figure}

For a number of different values of slopes $\gamma$
and length scales $\alpha$ we constructed a series of two-dimensional
images from a model disk of the corresponding profile, seen under the
inclination of M\,94 ($i = 35^{\circ}$ determined from the surface
photometry of the inner disk, see Sect.~\ref{isoph}).
For this modelling we chosed a high pixel-resolution of 1 pix $\cor$
0.25\arcsec. These model images were then convolved with a Gaussian
seeing profile of FWHM = 1.76{\arcsec}. Then an ellipse fit was
applied to every model image. In this way we obtained a two-parametric
manifold of curves ($\gamma = 0.7, 0.8, ... 1.7$ and $\alpha = 12, 15,
17, 20, 25, 30$\arcsec) which we compared with the observed ellipse
fit curve.  Fig.~\ref{figgam} shows the profiles of a $\gamma$--series
for $\alpha = 15${\arcsec} against $\log\,r$ compared to the
observation. The best fit for the inner region $r \le$ 30{\arcsec} $-$
40{\arcsec} was the curve for $\gamma = 0.9\pm0.05$ and for $\alpha = 15
\pm1.5${\arcsec}. This is very similar to Hernquist's (1990) luminosity
profile, which corresponds to $\gamma = 1$. For the
outer part of the model disk we added a simple exponential profile
with $r_{e} = 60${\arcsec} and neglected the oval shape of the outer
disk. This is justified since our model disk is intended mainly for
the inner region of M\,94, near and inside the inner ring.

%--------------------------------------------------------------------------
\subsection{Distibution of the optically emitting gas}
\label{emiss}

\begin{figure}
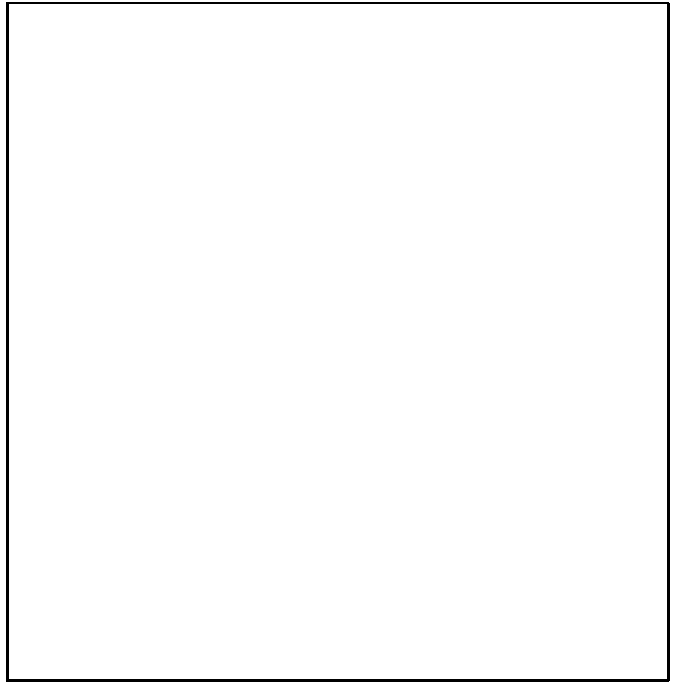

\picplace{9.0cm}
\caption[]{
$H_{\alpha}$ image of M\,94, $2 \times 2$ arcmin, north is up.
The continuum was subtracted using an
appropriate red image. Emission of warm gas is present in the
ring and in a central region which coincides roughly with the bar.}
\label{figalph}
\end{figure}

What is the distribution of the warm ionized gas in M\,94? For that
purpose a CCD image with an $H_{\alpha}$ filter was exposed. Since
this image still contains a strong continuum contribution, we used the
$I$ frame as our most appropriate continuum image. Since both images
have different pixel sizes, we rebinned the $H_{\alpha}$ image ($27
\mu$ pixels) to the pixel size of the $I$ image ($22 \mu$). Then both
images were centered to $\sim0.1$ pix and were subtracted with an
appropriate intensity scaling which made the pure continuum regions
vanish. Fig.~\ref{figalph} shows the result, it is very similar to
Fig. 4d in Pogge (1989). The main $H_{\alpha}$ contribution comes from
the well known ring of star forming $HII$ regions. There exists also a
central component of irregular, elongated shape; this extends mainly
parallel to the stellar bar (compare also the $H_{\alpha}$ map in
Duric \& Dittmar, 1988). A weak and also irregular spiral structure
can be detected between the ends of the bar and the strong ring at $r
= 45$\arcsec. The defects in the very center are due to the different
seeing during $H_{\alpha}$ and continuum exposures.

%--------------------------------------------------------------------------
\subsection{Color index images}
\label{col}

\begin{figure}
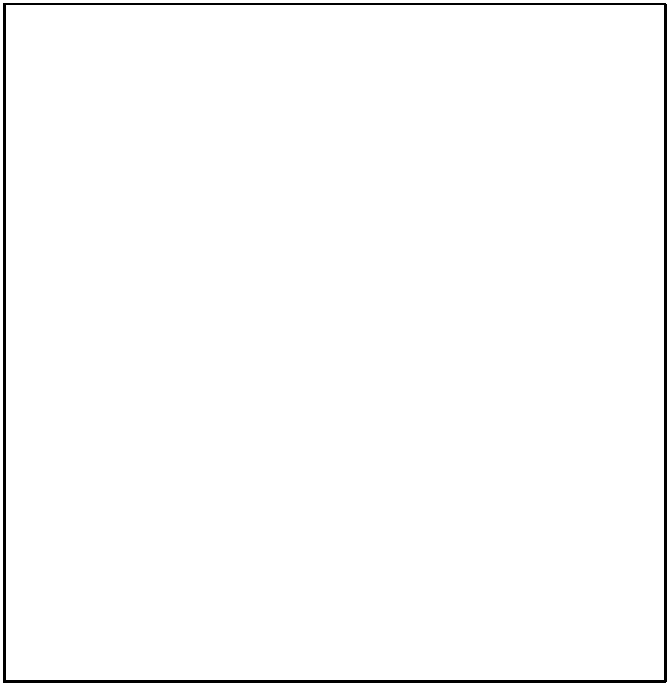

\picplace{9cm}
\caption[]{
Color index image $I - K$ of the central parts of NGC\,4736
(M\,94), $2 \times 2$ arcmin, north is up.
There exist some dust-reddened arc regions near the center
(dark shaded), however they are not responsible for the deformations
of the isophotes in the $I$ image. Long dust lanes of spiral arm shape
are crossing the ring of $HII$ regions at $r\approx45${\arcsec}.}
\label{figik}
\end{figure}

$V - I$, $I - K$, and $J - K$, color index images were produced to
study the morphologhy of dust absorption in M\,94.  The two
respective images were rebinned, if necessary, to the same pixel
scale.  Then they were adjusted in position to an accuracy of $<0.1$
pixels and divided in order to obtain a color index image. Dust
reddening is present in the dark shaded regions in the color
images. Fig.~\ref{figik} shows, as example, the $I - K$ image. There is
some dust present in circular arcs around the center, especially in
the NE and W. These features seem to be correlated with the faint
spiral arms around the central bulge. They can also be detected in the
$B - V$, $V - R$ images in Beckman et al (1991, Fig. 7a,b). Further out
two spiral-arm-like dust features follow the ring of $HII$ regions. All
these dust structures can already be seen in the $B$ broad band image
in Beckman et al (1991, Fig. 8). The most important point here is that
the inner dust arcs are situated near the long axis of the bar; they
would not be able to produce the bar feature by an adequate extinction
of a non-barred central stellar distribution in M\,94.

%--------------------------------------------------------------------------
\subsection{Stellar kinematics}
\label{reskin}

The CCD images of the stellar absorption spectra (4500--5400 \AA) were
reduced using the standard procedure with bias subtraction and
flatfield division. A special mask function was constructed from a
series of many differently exposed dome flats in order to correct for
the intensity-dependent additive column offsets of the RCA chip at low
light levels. The light distribution along the slit was corrected to a
flat profile by means of an apropiate function obtained from skyflat
exposures. The images were filtered to remove cosmics and then
rebinned row by row in $\log \lambda$ (equidistant in ${\Delta}v$).
The spectra were centered and placed exactly parallel to the
rows. During observation we had obtained two spectrograms for each
slit position. Now the reduced spectrograms of each pair were coadded
with an algorithm which removed further defects and remaining
cosmics. The continuum was subtracted by applying a polynomial fit to
every row. The $H_{\beta}$-- and $[OIII]$--lines were masked to avoid
disturbances by the emission lines. Then the spectrograms were
prepared for the cross-correlation by applying a $cos$-bell funtion to
their ends.  A similar technique was used for the preparation of the
template spectra of several G8\,III, K0\,III, K3\,III, and K5\,III
stars.

The stellar rotation velocity and the velocity dispersion of the
galaxy were determined by a Fourier-cross-correlation technique with a
standard template star; it is described in detail in Bender
(1990). This technique allows one also to obtain the line-of-sight
velocity distribution of the stars.

\begin{figure*}
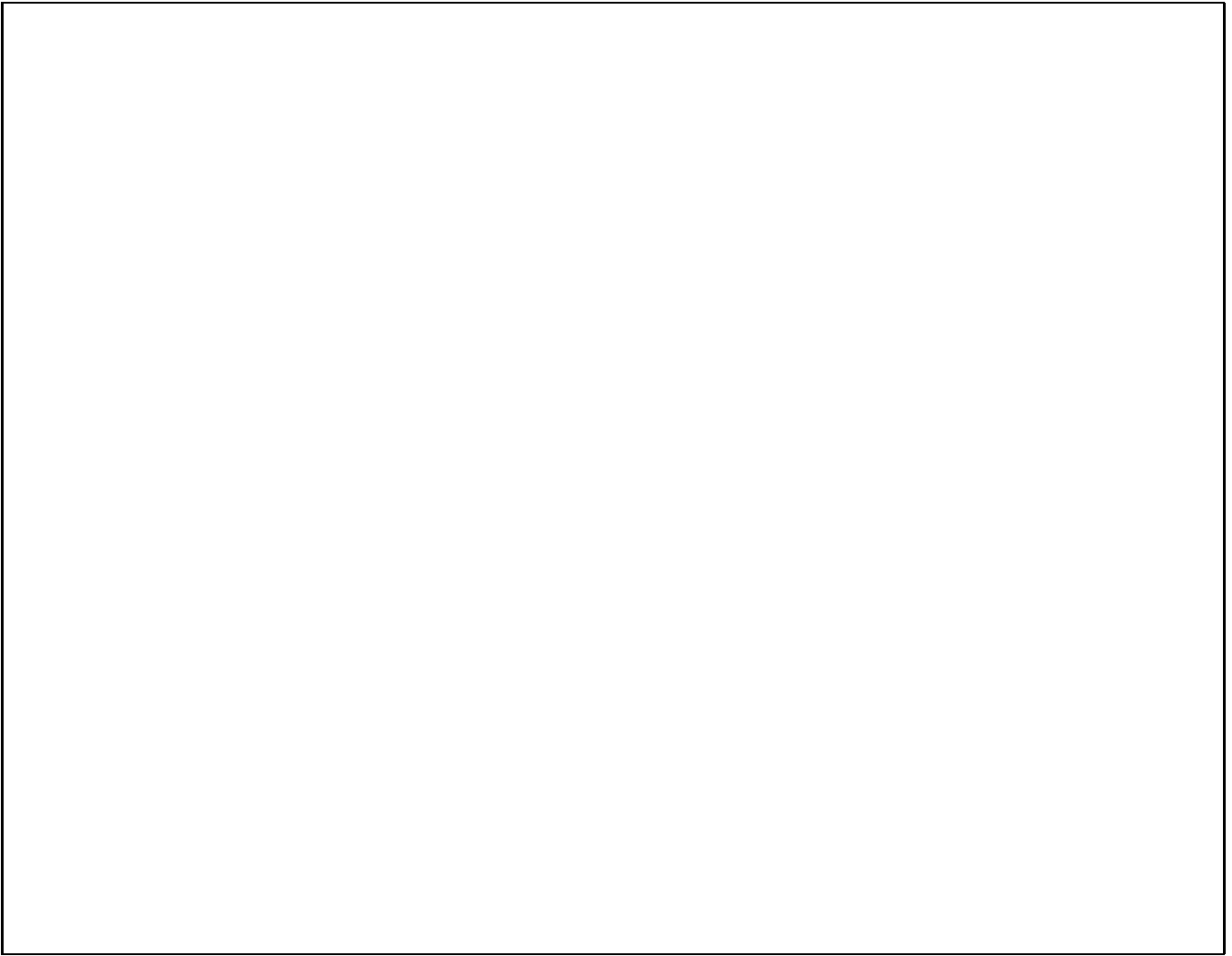

\picplace{14.0cm}
\caption[]{
M\,94: Stellar velocity dispersion (top) and stellar rotational
velocity (bottom) along $P.A. = 0^{\circ}, 45^{\circ}, 90^{\circ},
135^{\circ}$. The kinematical influence of the small bulge extends
only to $r \approx 20{\arcsec}\cor 600$pc.}
\label{figkin}
\end{figure*}

Fig.~\ref{figkin} shows the resulting stellar rotation (lower panel)
and stellar velocity dispersion (upper panel) along the four slit
orientations $P.A. = 0^{\circ}, 45^{\circ}, 90^{\circ},
135^{\circ}$. The strongest rotation amplitudes are seen along $P.A. =
90^{\circ}$ and $135^{\circ}$. Since the latter amplitudes are nearly
equal, the kinematic line of nodes must be at $\sim113^{\circ}$ and the
rotation axis at $P.A. \approx 23^{\circ}$. The stellar rotation
velocity increases nearly linearly to $r = 12${\arcsec}and then remains
on that level until $r \approx 60$\arcsec. For larger radii our signal
is too weak for a reliable result from the cross-correlation analysis.
The curves become noisy for $r > 40$\arcsec, this is due
to template mismatching in the ringlike zone of star formation.

Under the assumption of circular symmetry the stellar velocity field
can be easily deprojected. If this assumption is justified,
and all 4 rotation curves are deprojected
with the same correct inclination with respect to the correct line of nodes,
than the deprojected rotation velocity curves should fall together.
If the adopted inclination and line
of nodes are chooesen incorrectly, the projected curves are different.
In this way we determined by iteration $i=30^{\circ}\pm 5^{\circ}$ for
the inclination, and a position angle of $P.A. = 113^{\circ}\pm1^{\circ}$
for the line of nodes. The deprojected rotation curves coincide
quite well, especially outside the bar.
{}From the superposed projected curves it is possible to
calculate a mean stellar rotation curve with a better
signal-to-noise-ratio (see next Section). This 'kinematic' inclination
angle is very similar to that obtained by surface photometry of the inner
region of the disk ($i=35^{\circ}$ between $r=50$ and 100{\as}, see
Sect.~\ref{isoph}).

In the stellar velocity dispersion profile (upper panels of
Fig.~\ref{figkin}) we see a rather sudden increase inside $r=20${\arcsec} to
the central
value of $\sigma_{o} = 120$ \kms. This is most
plausibly interpreted as the signature of a small central bulge with
$r\approx20\as$ in M\,94.  Outside the bulge we obtain $\sigma = 60$
\kms, a value which is determined by our spectral resolution and is
probably higher than the velocity dispersion of the disk. The extent
of the bulge is fairly independent of the slit orientation, therefore
its shape is nearly axially symmetric.

The broadening function of the stellar absorbtion lines in a galaxy
spectrum is produced by the local velocity distribution integrated
along the line of sight.  The broadening function is gaussian only for
simple dynamic systems; in general it is a more complex function,
which we call the line-of-sight stellar velocity profile (LOS-SVP).
we use the method of Bender (1990) to determine this function by a
deconvolution of the correlation peaks.  Fig.~\ref{figlos} shows the
line-of-sight stellar velocity profiles of the spectra along $P.A. =
90^{\circ}$ at different radial distances from the center (0\arcsec,
$\pm5$\arcsec, $\pm10$\arcsec, $\pm20$\arcsec). While the central
profile is broad and gaussian (bulge profile), the line profiles at
$\pm5${\arcsec} are already asymmetric. At $\pm10${\arcsec} the
profiles are more complex: we see a narrow peak from the rapidly
rotating disk-stars, superposed onto a broader, more slowly rotating
component from the bulge stars.
At this radius the contributions of disk and bulge are
approximately of the same magnitude.  At $\pm20${\arcsec} only the disk
profiles are visible on both sides.

These line-of-sight velocity profiles confirm the existence of a bulge of
$\sim20${\as} radius in M\,94. Kormendy (1993) used the old
kinematical data of Pellet \& Simien (1982) to argue from
$v_{rot}/\sigma \approx 0.75$ and $\varepsilon \approx 0.1$ that M\,94
has no bulge but is disklike even in the center. This argument is no
longer valid since we now know that it compares velocities from
two different stellar
components.  From the velocity decomposition in Fig.~\ref{figlos} we
see that at $r=\pm10${\arcsec} the bulge component has a much smaller
rotation velocity than the disk component; thus $v_{rot}/\sigma$ for
the bulge stars becomes much smaller, and is not in conflict with the
classical $v_{rot}/\sigma - \varepsilon$ - diagram. Furthermore it
should be emphasized that spiral structure is not detectable inside
the bulge radius (20{\as}) of M\,94 (e.g. Fig.~\ref{figik}).

\begin{figure}
\picplace{9.0cm}
\caption[]{
Line-of-sight stellar velocity profiles (LOS-SVP) along $P.A. =
90^{\circ}$ at different radial distances from the center (0\arcsec,
$\pm5$\arcsec, $\pm10$\arcsec, $\pm20$\arcsec). While the central
profile is broad and gaussian (bulge profile), the influence of the
bulge decreases with distance from the center.  The combined profile
at $\pm10\as$ shows the broad, non-shifted bulge profile and the
rotationally shifted narrow disk profile. At $\pm20\as$ only the
narrow disk profile can be seen.}
\label{figlos}
\end{figure}

%----------------------------------------------------------------------
\subsection{Emission gas kinematics}
\label{resgas}

The reduction of the emission line spectra was performed in a very
similar way as described above for the stellar spectra. In order to
extract the emission lines from the stellar continuum, the latter was
subtracted via a polynomial fit (with masked emission lines) in every
row. The velocity field of the gas is then easily measurable in
the $H_{\alpha}$, $[N II]$, and $[S II]$ emission lines.
The radial velocity of the gas was determined by cross-correlation
with the emission line spectrum from the galaxy center.  Since there
the $H_{\alpha}$ emission is compensated by the absorbtion of the
stellar continua, we also used as templates spectrograms from the star forming
rings (and correcting for their rotational offset).

\begin{figure*}
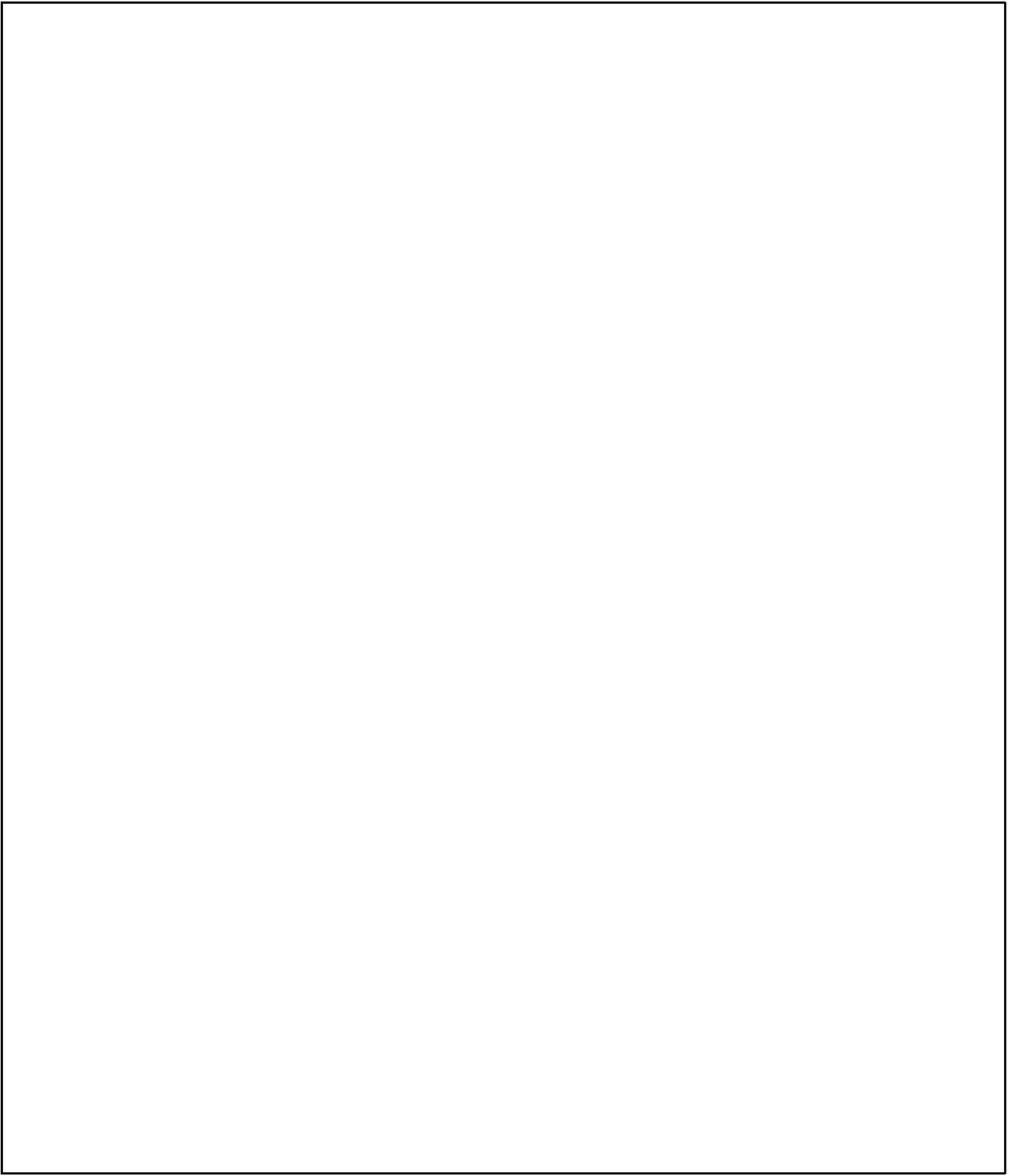

\picplace{21.0cm}
\caption[]{
M\,94: Rotational velocity of the gas (circles) along 8 different
slit positions, marked in each panel. The bulk rotation of the gas is
similar to the stars (solid line).  However, the gas rotates faster
and shows additional humps and dips in the curves.}
\label{figking}
\end{figure*}

Fig.~\ref{figking} shows the resulting gas rotation curves for all 8
observed slit orientations.  To compare the gas velocities with the
stellar rotation field we use the mean stellar rotation curve from the
last Section and project it along the corresponding slit position
angles. This results in the solid lines in each panel of
Fig.~\ref{figking}.  The rotation curves of gas and stars show a
similar general trend, i.e. practically disk dominated rotation for both
components, however, with some marked differences. They are best studied
by subtraction of the mean {\it stellar} rotation from the gas velocity
curves for each slit orientation.

(1) Inspection of the velocity residuals shows that the gas velocities
are systematically higher than the stellar mean velocities.  These
systematic differences are not an artefact of our mean stellar
rotation curve, they can be seen directly by comparing measurements of
gas and stellar rotation for the common position angles $P.A. =
0^{\circ}, 45^{\circ}, 90^{\circ}$, and $135^{\circ}$. Several
interpretations are possible.

(i) The gas rotates intrinsically faster than the stars.

(ii) A second possibility could be a different rotation axis for stars
and gas.  As we have seen in the last section, the best-fitting inclination
of the
stellar rotation plane is $30^\circ \pm5^\circ$.  A simple increase
of the inclination of the gas orbits to $\approx45^{\circ}$ and a
slight change of the line of nodes could explain the general trend
(assuming circular gas orbits), but this would mean that the gas
rotates in an inclined orbit with respect to the stellar rotation
which is a rather unplausible suggestion.

(iii) A third possibility is that the gas moves on elliptic orbits
along an axis roughly parallel to the small central bar. Deviation of
the gas orbits from circular motion were already discussed by van der
Kruit (1976), when he measured the gas velocities in the
$H_{\alpha}$-ring. Strong deviations from the pure circular motion of
the $HI$ gas were also observed by Mulder \& van Driel (1993) in the
form of variations in inclination and the line of nodes of the HI
velocity field. Our observed maximal amplitude of 120 km/sec for the
gas velocities at $P.A. = 118^{\circ}$ between $r = 20${\arcsec} and
40{\arcsec} matches well to the corresponding $HI$ velocities in Mulder \&
van Driel (they use $i = 40^{\circ}$ in their Fig. 12b).

A further plausible explanation of the systematic differences
between stellar and gas rotation is given in Sect. 5.

\begin{figure}
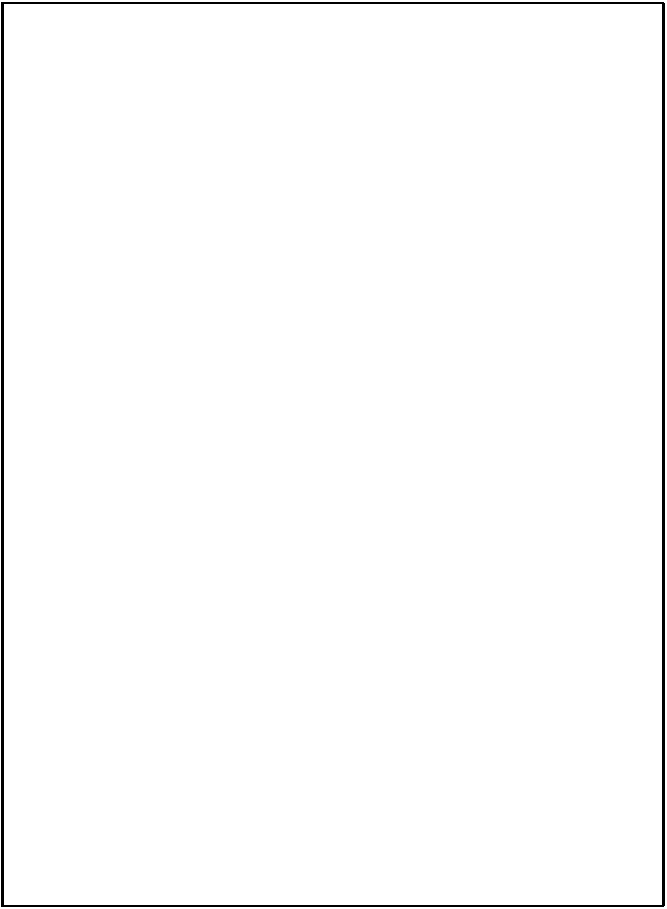

\picplace{12.0cm}
\caption[]{
M\,94: Locations of the humps and dips in the gas rotational
velocities relative to the mean stellar rotational velocities along 8
different slit positions. Open circles refer to negative, filled
circles to positive residual velocities, their diameter gives the
absolute values. The deviations concentrate along circular arcs, probably
connected to the inner spiral structure.}
\label{fighumps}
\end{figure}

(2) In the gas rotation curves there appear 'overshooting' and
'counter-rotating' humps and dips of the gas velocity near the center
of the galaxy. Fig.~\ref{fighumps} shows the location of these local
deviations by filled (humps) and open (dips) circles.
Since many of them occur at similar radii (10{\arcsec} to
20\arcsec) for different position angles, we conclude that the reasons
for these local deviations are of nearly circular shapes, probably spiral
arms. The eastern arc of filled cirles (redshifted deviations
with respect to the general velocity field) can roughly be identified
with the corresponding dust arcs displayed in the $I - K$ image
(Fig.~\ref{figik}). However, there also seems to be some connection
with the bar: The deviations occur mainly along the long sides of the
bar and they are mostly positve on the eastern side and negative on
the western side. In contrast to the general trend these deviations
can not be eliminated by a varition of the inclination angle and/or
the line of nodes.

A comparison with the dynamical model will be given in
Sect.~\ref{modkin6}, the discussion of the results in
Sect.~\ref{discuss}.

%----------------------------------------------------------------------
%----------------------------------------------------------------------
\section{Morphological model: Central bar or triaxial bulge?}
\label{modmorph}

In Sect.~\ref{resccd} we described the isophote twist, the variations
of the ellipticity, and the occurrence of cusped isophotes in M\,94
inside 40{\arcsec}(Fig.~\ref{figi} and Fig.~\ref{figell}). It could be
shown from the $I - K$ color index image that these are not an effect
of dust absorption.  Are the isophote twists etc. the signature of a
bar which is a disk phenomenon, are they caused by a triaxial bulge,
i.e. a structure extended in z-direction?

For the construction of a morphological model of the inner
region of M\,94 we have to consider the following observational facts:

a) isophote twist of nearly $90^{\circ}$ between 15{\arcsec}and 25\arcsec

b) round isophotes inside 10{\arcsec}which can not be explained just by seeing

c) cusped isophote shapes at $r\approx15\arcsec$

d) increased stellar velocity dispersion inside 20\arcsec, independent
   of the position angle.

We present two different models for the interpretation of these
observational findings: (1) M\,94 consists of a disk and an
axisymmetric (spheroidal) bulge, the isophote twist and the cusped
shapes of the isophotes are due to a weak additional bar component.
(2) M\,94 has a triaxial bulge in the center which is responsible for
the isophote twist (no direct explanation for the cusped isophote
deviations).

%----------------------------------------------------------------------
\subsection{Axially symmetric bulge, disk, and bar}
\label{modmorph1}

This model with an axisymmetric bulge is supported by the fact that
the stellar velocity dispersion profile is identical for 3 of 4 slit
orientations. The $\sigma$ - profile along $P.A. = 135^{\circ}$ is somewhat
broader; however, we do not consider this as significant.

Since we are mainly interested in the structure of the inner kpc, we
assume that the entire disk of M\,94 is axisymmetric and neglect its outer
oval shape. The modelling proceeded as follows:
According to the
isophote ellipse fits a seeing-convolved $\gamma = 0.9$
(Dehnen \& Gerhard 1994)
model for the inner disk (plus bulge), and an outer exponential disk
component were constructed. This two-component model together with an
additional constant value for the sky background fitted the whole disk
of M\,94 very well. The model parameters for the inner disk were:
$\gamma = 0.9$, length scale $\alpha = 15$\arcsec, and from the
surface photometry of the inner disk region (Sect.~\ref{isoph}):
inclination angle $i = 35^{\circ}$, position angle $P.A. =
113^{\circ}$.

This model was convolved with a Gaussian of FWHM =
1.76{\arcsec} according to the seeing during observation (measured
from the point spread function of stars in the $I$-image). For the
outer disk we adopted an exponential profile with $r_{e} = 60$\arcsec,
with inclination and position angle as above. The central intensity
was adjusted to the $I$ image.

\begin{figure}
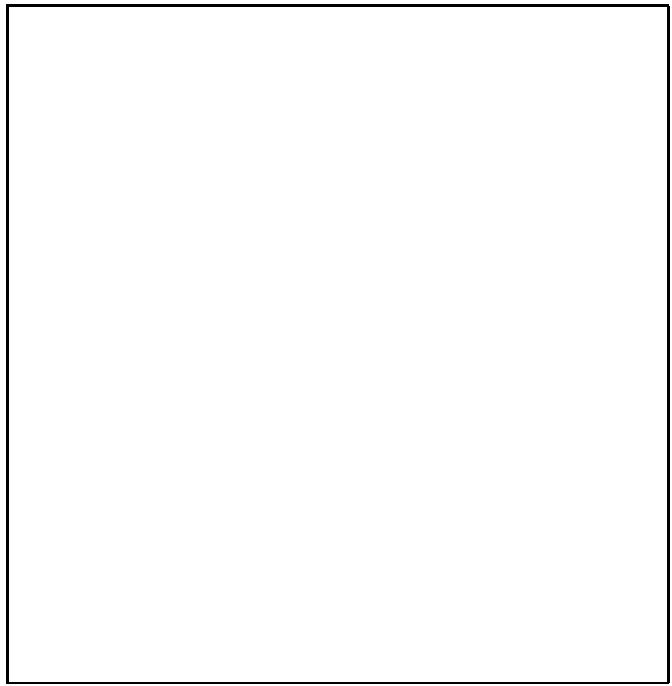

\picplace{9.0cm}
\caption[]{
Residual central bar component in M\,94 after the subtraction of a
two-component model for bulge ($\gamma = 0.9)$ and exponential
disk. The size of the bar is $30 \times 14${\arcsec} in projection.
The scale is the same as
the $I$ image in Fig.~\ref{figi} ($78 \times 78${\arcsec}, north is up.}
\label{figbar}
\end{figure}

When this model was subtracted from the observed $I$ image, a central
barred component remained, with a fairly flat
profile. Figure~\ref{figbar} shows the residual central bar. The
size of the bar is about $30 \times 14${\arcsec} in
projection. At the distance of NGC\,4736 ($D = 6.6$ Mpc,
$1{\arcsec}\cor 32 pc$) and for an inclination of $i = 30^{\circ}$, the
size of the bar from end to end is $42 \times 14\as$ or $1.3 \times 0.45$ kpc.

In a second step the central bar itself was modelled. As Fig.~\ref{figbar}
shows, the luminosity profile along the bar is rather shallow in the center and
steep at the boundaries. Therefore an elliptical $r^{1/4}$ profile is not
adequate. A flatter Ferrers profile (Ferrers 1877) was chosen. It has the
following elliptical surface brightness distribution:

\begin{equation}
   \Sigma(m^2) = \left\{
                     \begin{array}{ll}
                        \Sigma_0 (1-m^2)^n,\ \  &\mbox{if $m^2<1$},\\
                        0,                      &\mbox{else, }
                     \end{array}
                   \right.
\label{eq-ferr-dens1}
\end{equation}
\noindent
where

\begin{equation}
   m^2 = \left(\frac{x}{a}\right)^2
		      + \left(\frac{y}{b}\right)^2.
\label{eq-ferr-dens2}
\end{equation}
\noindent
The parameter $n$ determines the radial density profile, $\Sigma_0$ is
the central surface brightness. The long and short axes of the bar are
denoted by $a$ and $b$, respectively. For our model we chose $n = 2$,
$a =15\as$ and $b = 7\as$ (in projection).

\begin{figure}
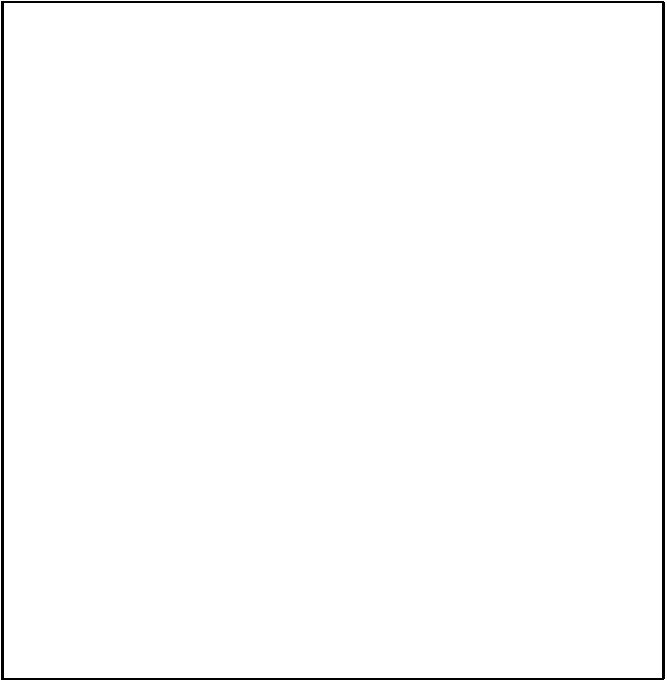

\picplace{9cm}
\caption[]{
Two-component model for the disk and bulge together with a Ferrers
model for the central bar component in M\,94. Compare this composite
model with the $I$ image in Fig.~\ref{figi}; the same scale and lookup
table were used.}
\label{figpmod}
\end{figure}

Fig.~\ref{figpmod} shows the coadded model disk and Ferrers bar.
Subtraction of this composite model from the $I$ image left only
very small residuals.

%----------------------------------------------------------------------
\subsection{Triaxial bulge and disk}
\label{modmorph2}

One of the main characteristics of the above model is the fact that
the overwhelming portion of the mass in the central region is contained in
the axisymmetric component, which results in a dominant monopole
component of the potential in the disk plane.
Compared to this the quadrupole component
due to the bar is only a minor perturbation of the order of a few percent.
If the light in the inner 10{\as} is dominated by a single triaxial
component, this can therefore only have a small flattening in the disk
plane of M\,94.

We have the following observational boundary conditions for the construction
of a triaxial bulge:

a) The long axis of the bulge is along $P.A. = 28^{\circ}$.

b) From the line-of-sight-velocity decomposition (Fig.~\ref{figlos}) we
   conclude that the amount of light from the bulge at $r\approx10\arcsec$
   is approximately equal to that of the disk there.

c) From the $\sigma$ - profile (Fig.~\ref{figkin}) the bulge light inside
   $\sim10\arcsec$ is dominant compared to the disk light.

The construction proceeded as follows: The azimuthally averaged luminosity
profile from
the ellipse fit to the observed $I$ image (Fig~\ref{figell}) was
fitted by a two-component model including a steeper $\gamma$ - profile
for the bulge and a flatter $\gamma$ - profile for
the disk. The contribution of these profiles had to be equal at
$r\approx10\arcsec$, and their sum had to fit the observed total
profile. A good solution was found with $\gamma = 0.9$ and $\alpha =
15\arcsec$ for the bulge component, $\gamma = 0.8$ and $\alpha =
205\arcsec$ for the disk component, with a ratio of their total light
B/D = 1.044. It is obvious that these two components have a rather
similar structure.  Furthermore this decomposition is not unique and
the astrophysical meaning of the labels 'bulge' and 'disk' for
these components is not clear.

Using these profiles we constructed the projected image of an
elliptical prolate bulge with $b/a \approx 0.3$ and its long axis
along $P.A. = 28^{\circ}$ and an inclined disk ($i = 35^{\circ}$, node
line along $P.A. = 113^{\circ}$) and coadded their projection as a
twodimensional model image of M\,94. The result was not a good
description for M\,94, since the bulge component was too dominant to
give the correct isophote twist at the correct radius. The same is true with a
centrally flatter disk profile.  Vice versa, if we weakened the bulge
component or made its profile steeper, we could not satisfy above
conditions b) and c). A number of different bulge axis ratios
and bulge-disk combinations were tried. The most serious difficulty
is that a bulge with a density profile as above has no sharp
boundary, even though the bulge profile steepens to $r^{-4}$ for $r >>
15\as$.  Therefore no cusped isophotes are produced. A bulge does not
have a natural outer radius like a bar at its corotation radius.  Of
course, it would be possible to use more complex components, e.g. a
triaxial bulge with changing axis ratios or a bulge which is not
elliptical or which has a sharp outer boundary. However, then too many
arbitrary parameters would come into play and the result would not be very
meaningful.

We conclude that our first model is the most plausible: M\,94
consists of a disk with a Hernquist like profile ($\gamma \approx 1$) and a
spheroidal bulge which can only be separated from the disk by its stellar
velocity dispersion but not by its light profile or color. The deviations and
twists of the isophotes are due to a superposed weak bar component.
The bar ends inside the gas ring at its corotation radius.

%----------------------------------------------------------------------
%----------------------------------------------------------------------
\section{Dynamical model}
\label{modkin}

In this section we present a  model for the kinematics in the inner region
of M\,94.  We will try to explain the radial velocity curves of the
stars and of the gas emission, especially of the 'overshooting' and
'counter-rotating' components (humps and dips, Fig.~\ref{figkin}).

As was shown in Sect.~\ref{reskin} the rotation curves of stars and
gas are qualitatively similar.  The velocity dispersion of the stars
outside $r=20$\arcsec is below 60 \kms, which is
the observational limit. So we can assume that the kinematics of the
stars outside the bulge region is disk-like.  For the bulge we have a
projected stellar
rotation velocity $v_{rot} \approx 50$ \kms and a velocity dispersion
$\sigma \approx 120$ \kms (Fig.~\ref{figlos}).  From the well known
relations between $\varepsilon$ and $v_{rot}/\sigma$ we estimate a true
flattening of $\epsilon\approx 0.4$ for the bulge.

%----------------------------------------------------------------------
\subsection{Mass model}
\label{modkin1}

We construct our mass model according to the morphological
components from Sect.~\ref{modmorph1}:

\begin{itemize}

\item The bulge and inner disk is approximated by a flattened
$\gamma$~-~model with $\gamma=0.9$, $\alpha=15\as$, and $q=0.6$.

\item The bar component is approximated by a Ferrers~bar:
$a=20.8$\arcsec, $b=7.3$\arcsec, $n=2$.

\item The contribution of the outer exponential disk
for the mass model of the inner bar region is neglected.

\end{itemize}

The mass-to-light-ratio $\zeta$ was determined by fitting circular
orbits at $r=45\as$ (inner ring) and was assumed to be equal for
both components (no systematic color gradient was observed).

The flattened $\gamma$ -- profile for the inner disk and bulge is:

\begin{equation}
\rho_q(p) = \frac M{2\pi q} \frac \alpha{p^\gamma}
            \frac 1{(p+\alpha)^{4-\gamma}},
\end{equation}
with $p^2=x^2+y^2+z^2/q^2$, scale-length $\alpha$, and
$M=2\times 10^{10}M_\odot$.

The mass of the bar can be calculated by
integrating the bar surface brightness (Eq.~\ref{eq-ferr-dens1})
over all space:

\begin{eqnarray}
M &=& \zeta\int\!\!\!\int \Sigma(m^2=\frac{x^2}{a^2}+\frac{y^2}{b^2})\,dx dy\\
  &=& 2 \pi \zeta\,\Sigma_0 ab  \int_0^1\! dr\ r (1-r^2)^n \\
  &=& \zeta\,\Sigma_0 \pi ab \frac{\Gamma(n+1)}{\Gamma(n+2)},
\label{eq-bar-mass}
\eeqy
\noindent
where the last integral was evaluated by using integral tables
(e.g. Bronstein \& Semendjajew, 1987, 1.1.3.4 \#39).

Using the values above one gets a bar mass of $9.8\times 10^8
M_\odot$, which is about 14\% of the mass of the flattened $\gamma$ --
model up to the end of the bar ($r = 20${\arcsec} in the plane of the
galaxy, $7.0\times 10^9 M_\odot$).  From the surface photometry
calibration we obtain $I=7.8$ inside that circle.  With the color indices
of de Vaucouleurs \& Longo (1988) for the inner 35{\as} aperture
of M\,94, we obtain $B=10.6$ inside of $r=20\as$, and with a distance of
6.6 Mpc this leads to $M/L_B = 1.8$. This extraordinary low value was
already obtained by Buta (1988); it is a consequence of the extremely high
surface brightness of M\,94.  Fig.~\ref{fig_density} shows the radial
surface density of the different components of our model.

The bar rotates rigidly with a
constant pattern speed $\Omega_b$, which is determined in Sect.~\ref{modkin4}.

\begin{figure}[hbt]
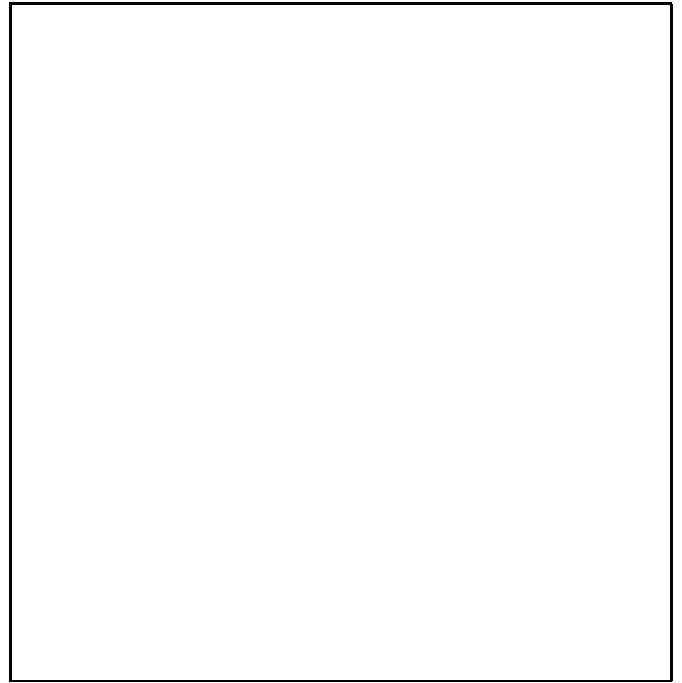

\picplace{9.0cm}
\caption[]{
Surface brightness of $\gamma$ - model, Ferrers bar and their sum
along the bar major axis in units of the characteristic length $\alpha$.}
\label{fig_density}
\end{figure}

%----------------------------------------------------------------------
\subsection{Gravitational potential}
\label{modkin2}

To calculate the families of stable closed orbits one needs the
potential of the galaxy model, from which one can determine the forces
exerted on a test particle. The potential of the flattened $\gamma$
-- model (Dehnen \& Gerhard 1994) is
\begin{equation}
 \Phi_q(R,z) = - \frac 12 \int_0^\infty \frac {\tilde\psi(\tilde m) d\tau}
					{(\tau+1)\sqrt{\tau+q^2}},
\end {equation}
where
\begin{equation}
 \tilde m = \sqrt{\frac{R^2}{\tau+1}+ \frac{z^2}{t+q^2}}
\end{equation}
and
\begin{equation}
 \tilde\psi(\tilde m) = \frac 1{2-\gamma}
 \left[
	1-\frac{\tilde m^{3-\gamma} + (3-\gamma) \tilde m^{2-\gamma}}
		{(\tilde m+1)^{3-\gamma}}
 \right].
\end{equation}

This expression was numerically calculated using a routine by Walter Dehnen,
whereas the potential of the Ferrers~bar was calculated with the following
expansion method described by Binney \& Tremaine (1987):

\beq
 \Phi(r,\varphi) = \sum_{k=0}^\infty \sum_{m=-k}^k V_{km} Y_{km}
 	 (\arcsin(r), \varphi),
\eeq
\noindent where the expansion constants $V_{km}$ are given by
\beqy
  V_{km} &  = &-\pi^2 G g_{km}  \int_0^{2\pi} d \varphi'
        \nonumber  \\ & & \times  \int_0^{R(\varphi')}\Sigma(r',\varphi')\,
                    Y_{km}^*(\arcsin(r'), \varphi') \,r'\,dr'.
\eeqy
The functions $Y_{km}$ are the well known spherical harmonics, $G$ is Newton's
constant of gravity and the $g_{km}$ are constants. The variable $r$ is
normalized in such a way that the density distribution lies inside a circle of
radius 1. $R(\varphi')$ denotes the edge of the density distribution. The
equations give the potential in a disk inside this circle.
The expansion was truncated at $k=30$ and $|m|=8$;
such high values were necessary to approximate the potential of the
elongated bar in M\,94 accuratly enough (the axis ratio is 1:2.85).
For rounder bars smaller values of $k$ and $|m|$ are sufficient.

%----------------------------------------------------------------------
\subsection{Resonances and families of closed orbits}
\label{modkin3}

The families of closed orbits in barred
potentials are determined mainly by the loci of the Lindblad- and
{\cR}-resonances ({\Cs} \& Grosb\o l, 1989).
For weak bars linear theory can be applied.
Then the resonances can be calculated in terms of two natural frequencies:
\beq
  \Omega_0^2 = \left( \frac 1R \frac{d\Phi_0}{dR} \right)_{R_0}
\eeq
is the circular frequency of a star on a circular orbit with radius
$R_0$ in the axisymmetric part of the potential ($\Phi_0$).
Small radial displacements from this circular orbit lead to
radial oscillations with the epicycle frequency $\kappa_0$, where
\beq
 \kappa_0^2 = \left( \frac{d^2\Phi_0}{dR^2} + 3\Omega_0^2\right)_{R_0}.
\eeq
The frequencies $\Omega_0$ and $\kappa_0$ are calculated only from
the axisymmetric
part of the potential and are accurate only for weak bars.
The exact positions of the resonances in strong bars have to be determined
from the behaviour of the families of closed orbits (see next section).

Azimuthal displacements do not change the
circular orbit, so the azimuthal frequency is 0. The resonances arise between
the forcing frequency seen by the star, $m(\Omega_0-\Omega_b)$ (for
barlike potentials $m=2$), and the two natural frequencies $\kappa_0$ and $0$.

If\ $\Omega_0-\Omega_b = 0$,\ the {\it \CR} resonance (CR) appears,
be\-cau\-se the star moves as fast as the rotating bar.
If $m(\Omega_0-\Omega_b) = \pm \kappa_0$, the star encounters successive
crests  of the potential at a frequency that coincides with that of the natural
radial oscillations. An {\it Inner Lindblad\/} resonance (ILR) occurs if
the star overtakes the rotating bar ($'+'$-sign),
an {\it Outer Lindblad\/} resonance (OLR) otherwise ($'-'$-sign).

A {\it resonance-diagram} can be plotted (Fig.~\ref{fig_resdiag}) and the
loci of
the resonances can be determined by drawing a horizontal line at
$\Omega_b$. The small oscillations in the curves near $r=40\as$
are numerical artefacts:
To calculate $\kappa_0$ one has to evaluate the second derivative
of $\Phi_0$. This axisymmetric part of the potential consists of the
flattened $\gamma$ -- model and the axisymmetric part of the bar potential,
which is given in a series. Even if the potential is approximated quite
well, the second derivative still has oscillations, which are visible
in Fig.~\ref{fig_resdiag}.

\begin{figure}[hbt]
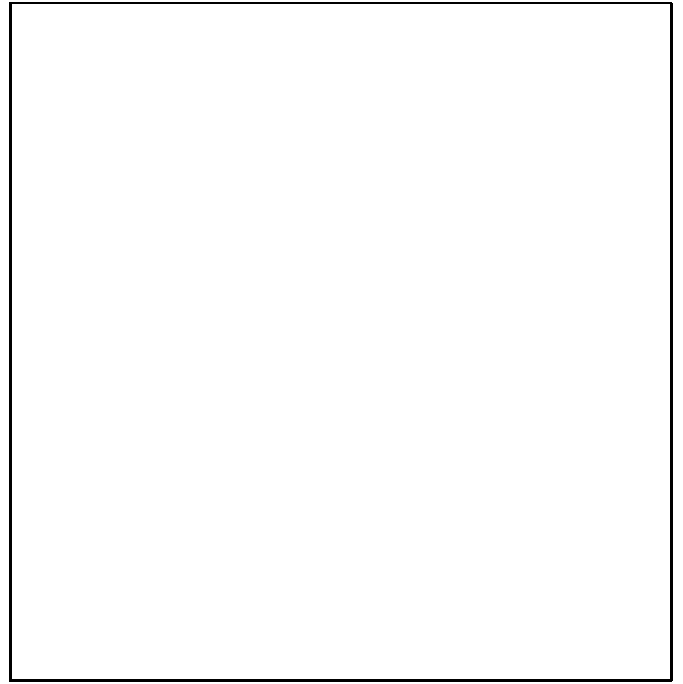

\picplace{9.0cm}
\caption[]{
Resonance diagram for our model. The horizontal line
$\Omega_b=2.5$ shows the resonances, which are marked below.
The bar ends at $r=20\as$. The oscillations in $\kappa$ are
due to the numerically calculated second derivative of the
potential with a limited order of the expansion.}
\label{fig_resdiag}
\end{figure}

The resonance diagram shows that because of the steep potential of M\,94
only one inner Lindblad resonance can occur.  In such a potential the
closed orbit families in the corotating frame of reference have the
following properties (see {\Cs} \& Grosb\o l, 1989 for a complete
review):

\begin{description}
\item{
The prograde $x_1$-orbits are elongated along the bar and exist from
the center outwards to nearly CR. These orbits should strongly sustain
the stellar bar shape.}

\item{
The prograde $x_2$-orbits are elongated perpendicular to the bar and
exist in the energy range all the way from the center
up to the energy of the ILR.}

\item{
The prograde $x_3$-orbits have the same properties as the $x_2$-orbits,
but are unstable, so they can not be populated with gas and are not
considered in the following.}

\item {
The retrograde $x_4$-orbits are not useful inside the CR, because they
would lead to a retrograde streaming of gas, which is not observed. }

\item{
However, outside the CR we have only retrograde orbits, which are nearly
round.  Inside the OLR they are slightly elongated perpendicular to the
bar, outside the OLR along the bar. These orbits can be populated with
gas, because a retrograde orbit outside CR has no retrograde
streaming in the rest frame. }
\end{description}

%----------------------------------------------------------------------
\subsection{Pattern speed}
\label{modkin4}

The pattern speed $\Omega_b$ is the only free parameter in our model
which can not be observed directly. Our aim was to determine $\Omega_b$
using the structure of the orbits of our mass model.

The stellar bar must be supported by $x_1$-orbits. As these exist only
inside CR, the pattern speed has to be so low that CR lies outside
the bar (Sellwood \& Wilkinson 1992).  Two other constraints on the
pattern speed are the location of the gas ring at $r\approx
45${\arcsec} and the gas-poor zone between $20${\arcsec} and
$40$\arcsec.  As was shown by Schwarz (1981) a rotating bar transports
gas from the area between CR and OLR to or just beyond the OLR. To
meet these two constraints, the pattern speed in our model was
determined at $\Omega_b = 2.5$ in our units, corresponding to one
revolution in $2.1\times 10^7$ years or 290 \kms per kpc. In this way
the OLR of the inner bar coincides with the ILR of the outer oval
disk: the star formation ring at $r \approx 45${\arcsec} (Gerin et al
1991).

%----------------------------------------------------------------------
\subsection{Kinematics of the stars}
\label{modkin5}

\def\voline#1{\overline{v_{#1}^2}}
How do the observed stellar rotation velocity and the velocity
dispersion in the bulge correspond to the observed mass distribution?
The radial Jeans equation for an axisymmetric stationary system in the
equatorial plane
\beqy
-\rho \frac{\partial\Phi}{\partial R} &=& \frac{\rho \voline R}{R}
 \left[
	\left(1-\frac{\voline \phi}{\voline R} \right)
	+\frac{ \partial \ln \rho}{\partial \ln R}\right. + \nonumber\\
& & \left.\qquad\quad	+\frac{ \partial \ln \voline R}{\partial \ln R}
	+ \frac R{\voline R} \frac{\partial \overline{v_R v_z}}{\partial z}
 \right]\quad (z=0)
\eeqy
(Binney, Tremaine 1987, p.197). Here $R$ denotes the cylindrical radius.
Assuming an isotropic rotator, we have
$\overline{v_R^2} = \overline{v_z^2} = \sigma^2$,
$\overline{v_R v_z} = 0$, and
$\overline{v_{\phi}^2} = v_{rot}^2 + \sigma^2$, and the last term in
the previous equation vanishes. Using
\beq
  \frac{\partial\Phi}{\partial R} = \frac{v_{circ}^2}{R}\quad(z=0),
\eeq
one gets
\beq
 v_{circ}^2 = v_{rot}^2 - \sigma^2
\left( \frac{\partial\ln\rho}{\partial\ln R}
        +   \frac{\partial\ln\sigma^2}{\partial \ln R}
\right),
\eeq
where $\sigma$ is the isotropic stellar velocity dispersion, $v_{rot}$ the
(observed) rotational stellar velocity, and $v_{circ}$ the circular velocity
in the $z=0$ plane in the potential corresponding to the density
distribution $\rho(R,z)$.
This equation is formally identically to that obtained from the Jeans equation
in spherical symmetry, but is only valid in the $z=0$ plane.

In this plane, using the density distribution of a Hernquist sphere
(Hernquist, 1990, corresponding to $\gamma = 1$)
\beq
  \rho(r) = \frac{Ma}{2\pi r}\frac{1}{(r+\alpha)^3}
\eeq
\noindent  we obtain
\beq
  v_{circ}^2 = v_{rot}^2 +\left( 4 - \frac{3\alpha}{R+\alpha}\right)\sigma^2,
\label{eq-vcirc}
\eeq
\noindent
neglecting the gradient of $\sigma^2$ compared to the gradient of the density.
The coefficient of $\sigma^2$ in this equation goes to 1 for $R
\rightarrow 0$ and to 5/2 for $R \rightarrow \alpha$.

The right hand side of Equ. (\ref{eq-vcirc}) is obtained from observational
quantities of the stellar velocity field. We take the mean stellar
rotation curve from Sect.~\ref{reskin} and deproject it using the
optimal inclination of $i = 30^{\circ}$.  We assume $\sigma$ to be
isotropic and spherically symmetric (no deprojection) and calculate
the mean velocity dispersion curve from the 4 independent stellar
spectrograms (Fig.~\ref{figkin}).  The corresponding values of
$v_{circ}^2$ in Eq.~\ref{eq-vcirc} are shown in Fig.~\ref{figjeans} as
circles.  These are compared with the circular velocity curve from the
axially symmetric part of the mass model (solid line in
Fig.~\ref{figjeans}). Since the M/L ratio is a free parameter
here, the amplitude of the circular velocity curve was adjusted to the observed
points. The important conclusion here is that observed points and the
circular velocity have the same radial profile; the stellar kinematics
confirm our mass model.

For comparison we plot in Fig.~\ref{figjeans}b the inclination-corrected
mean stellar rotation curve (+ symbols). This curve
obviously has a shallower gradient and a lower peak rotational
velocity, since it does not take into account the kinetic energy
in $\sigma$. We also plot the inclination corrected ($i =
30^{\circ}$) gas rotation curve at $P.A. = 118^{\circ}$ (triangles),
which is near to the line-of-nodes ($113^{\circ}$) The maximum gas rotation
velocity
reflects the mass model fairly well. We see that the gas rotates
faster than the stars, as expected; however, the central gradient is
again significantly shallower than that of the circular velocity
curve.

\begin{figure}
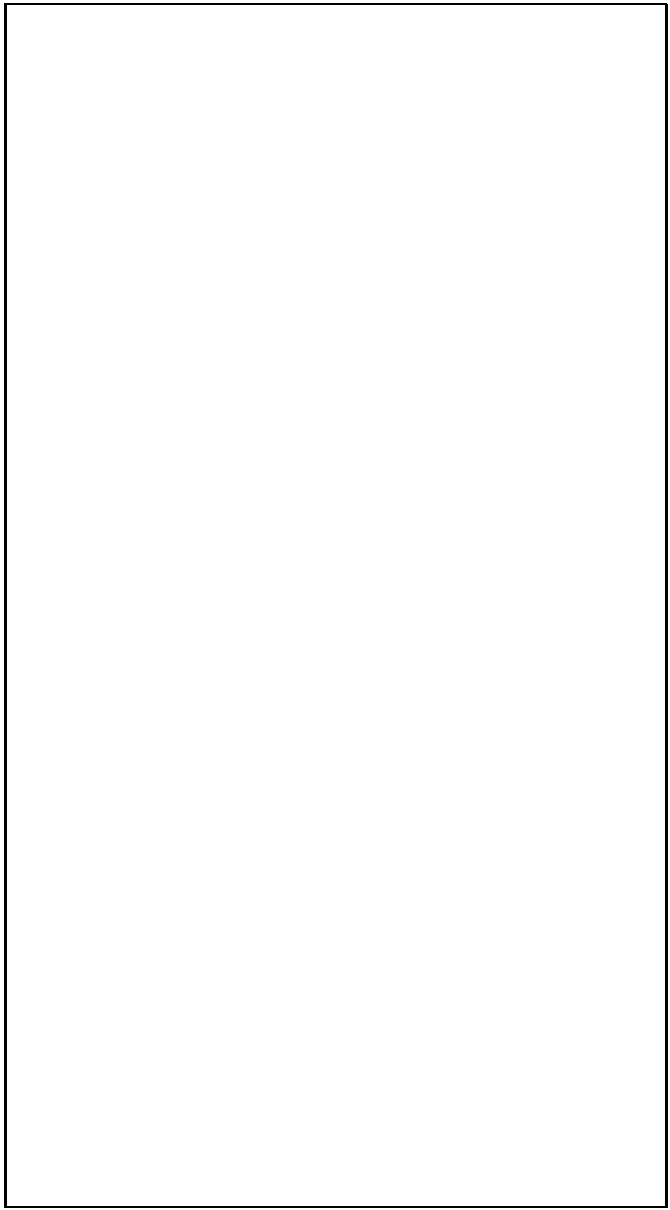

\picplace{16cm}
\caption[]{
{\bf a):} The circular velocity, as calculated from the observed stellar
rotation and velocity distribution by the Jeans equation (circles),
agrees well with the the circular velocity derived from the mass model in
Sect.~\ref{modkin1} (solid and dotted curve, resp.).
{\bf b):} Same as above, plus the stellar rotation
curve (+ symbols) and the gas velocity (triangles).}
\label{figjeans}
\end{figure}

%----------------------------------------------------------------------
\subsection{Kinematics of the gas}
\label{modkin6}

The mass model of Sect.~\ref{modkin1} was used to study the dynamics
of the optical line emitting gas.  We neglect the self-gravity of this
gas and consider it as test particles in the stellar potential.

Gas has a high cross-section, so encounters between gas clouds
normally have to be taken into account. These encounters lead to
dissipation of energy and angular momentum. As a first approximation
we assume that the gas clouds move on stable non-intersecting closed orbits
(Sellwood \& Wilkinson, 1993).

Figure~\ref{fig_orbits} shows the stable closed orbits in our model
(inside CR only prograde orbits are drawn).  The orbits are shown in
the corotating frame of reference.  All orbit integrations were done
using a 4th order Runge-Kutta integrator with variable time step and
error prediction. This was mainly taken from {\it Numerical Recipes in
C} (Press et al. 1992).

In the very center of Figure~\ref{fig_orbits} one member of the
$x_2$-family is plotted, which shows the small extent of this orbit
family. The ILR is at 5{\as} which is small compared to the scale of
the bar (20\as) and that of the figure.  The $x_1$-orbits are highly
elongated along the bar and become nearly round near the CR. Between
the CR and the OLR there exist no non-intersecting orbits, so gas is
swept out of this zone and piles up at the OLR.  Beyond the OLR there
exist again non-intersecting closed orbits that can be populated with
gas. These orbits are nearly round but slightly elongated along the
bar.

\begin{figure}[hbt]
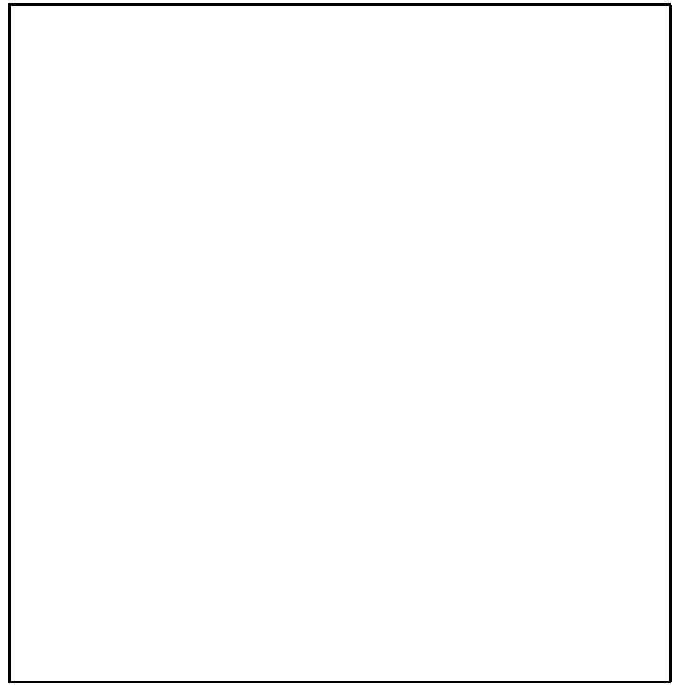

\picplace{9.0cm}
\caption[]{
Families of closed orbits in our model of M\,94. Resonances are
marked on the lower margin. While the center is populated by $x_1$-
and $x_2$-orbits, in the region between CR and OLR no stable
nonintersecting orbits can be found. Beyond OLR there exist nearly
round retrograde orbits (in the corotating frame) which make up the
ring.}
\label{fig_orbits}
\end{figure}

To obtain observable quantities from the orbits in
Figure~\ref{fig_orbits}, one has to transform their velocities from
the corotating frame to inertial coordinates and then project the orbits onto
the sky.  From the resulting velocity field we calculate rotation
curves at various position angles.  Here we have to take into account
that the velocity field is smoothed by atmospheric seeing
(1.5{\arcsec} FWHM), by the finite resolution of the spectrograph (120
km/sec FWHM), and by the finite slit width (4.3\arcsec).  Therefore
the gas velocity field has to be folded with the observed gas distribution
in the region of the galaxy contributing to the corresponding slit
position.  To obtain a model of the gas distribution, we consider an E-W
cut of the $H_{\alpha}$ distribution (Fig.~\ref{figalph}), this is
shown in Fig.~\ref{fig_gasring}. We approximate the central emission peak
by an axially symmetric distribution with a radial profile
$\sim (r+2{\arcsec})^{-1}$, and the ring at radius $r = 38$\arcsec
by a Gaussian profile (8{\arcsec} FWHM).

\begin{figure}[hbt]
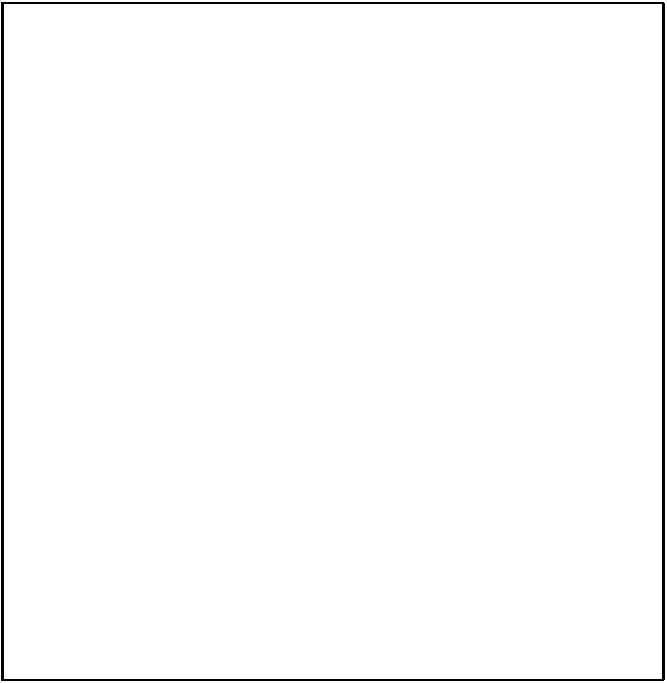

\picplace{9.0cm}
\caption[]{ Emission gas density along a cut through the center of M\,94.
Three peaks in the gas density are visible. The inner peak corresponds
to the center of the galaxy, while the outer ones at $45{\arcsec}$ are due to
the star forming ring, which is shown in Fig.~\ref{figalph}. The
resonances in our model are marked on the lower margin.}
\label{fig_gasring}
\end{figure}

Two example rotation curves were obtained by this procedure and are
shown by the $\times$ symbols in Fig.~\ref{fig_rotcurv}a,b. The
circles are the observed gas velocities along the corresponding slit
positions. The dotted curves give the minimum and maximum velocities
which occur in the considered slit area before smoothing by seeing and
finite resolution.

Fig.~\ref{fig_rotcurv}a shows the model rotation curves along
$P.A. = 118^\circ$, i.e. orthogonal to the bar.  At radii greater
$\approx 20${\arcsec} the agreement is quite good. This is because the
orbits are nearly round there, so the velocities of the orbits reach
the circular velocity, which was used to calculate the mass of the
model.  However, between 5{\arcsec} and 10{\arcsec} the model curve
shows a strong overshooting which arises from the contribution of the
$x_1$ orbits along the direction of their maximum streaming velocity.
The observed rotation curve is less steep in the center and does not
show such strong humps. This is similar for the adjacent slit
orientations (c.f. Fig.~\ref{figking}).

Fig.~\ref{fig_rotcurv}b shows the model rotation curves along
$P.A. = 28^\circ$, i.e. parallel to the bar.  As could be expected the
rotation curve along the bar is nearly flat since the bar axis is
nearly perpendicular to the line of nodes.  Our model does not show
the observed humps at this slit position.  However, the maximum and minimum
velocities are rather high, since our broad slit always covers the
innermost $x_1$ orbits in their entirety. The slit smoothing
gives the displayed smooth curve, but this may depend on the precise
distribution of gas that contributes to the observed velocities. So the
discrepancy in Fig.~\ref{fig_rotcurv}b is not as serious as in
Fig.~\ref{fig_rotcurv}a.

\begin{figure}[hbt]
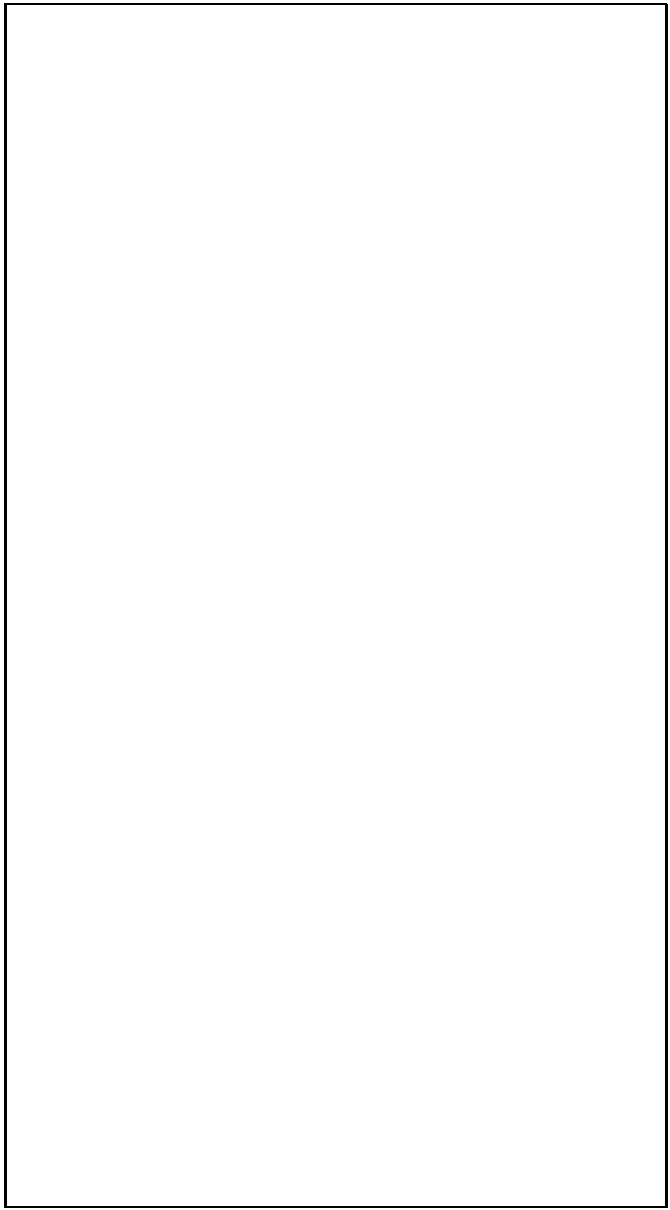

\picplace{16.0cm}
\caption[]{ Rotation curves along $\PA=118^\circ$ (top) and
$\PA=28^\circ$ (bottom). The circles are the observed velocities, the
crosses are the theoretical values.  The dots give the minimal and maximal
velocity in every seeing element.  For radii greater than $\approx
20${\arcsec} the agreement is quite good. At the center the rise of
the theoretical velocities is too steep to match the observations.}
\label{fig_rotcurv}
\end{figure}

Thus our kinematical model is not able to reproduce the observed
peculiarities of the gas rotation curves. From the model we expect
humps in the curves especially along the minor axis of the bar
(i.e. when the slit orientation is orthogonal to the bar) from the
$x_1$ orbits, and no humps on the major axis (spectrograms along the
bar).  Such a pattern is not observed, rather the distribution of
humps and dips follows more circular patterns (spiral arms,
Fig.~\ref{fighumps}).

As Fig.~\ref{figking} shows, the gas rotation velocities are somewhat
higher than the stellar velocities, but resemble these much better
than the curve in Fig.~\ref{fig_rotcurv}a. This is all the more
surprising since in the inner 15{\as} (the region of greatest
discrepancy) the bulge is hot and the kinematics in agreement with the
mass model (Fig.~\ref{figjeans}). This suggests that the $H_{\alpha}$
gas in the inner 15{\as} is both disturbed and still coupled to the
stars. Cold molecular gas might well follow the model velocity field
more closely,
but corresponding high resolution CO data unfortunately are not
available.

Our kinematical model was constructed in such a way that the outer
Lindblad resonance coincides with the gas ring at $r \approx 45$\arcsec.
Fig.~\ref{fig_ring_vel} shows the comparison of the observed gas velocity
on the ring compared with our model calculations.

\begin{figure}[hbt]
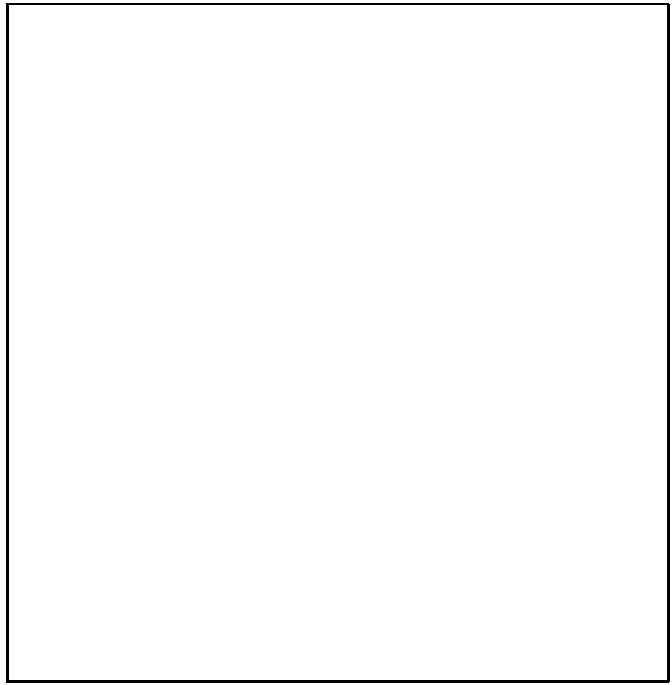

\picplace{9.0cm}
\caption[]{ Measured velocities of gas in the ring of M\,94
with respect to the center, as a function of position angle.  The
asterisks represent our measurements in Sect.~\ref{resgas}.  The
other symbols indicate different emission lines from van der Kruit
(1974).  The dotted lines give the theoretical predictions based on
closed orbits in the ring with approximate radii of $42\arcsec$,
$47\arcsec$ and $51\arcsec$.}
\label{fig_ring_vel}
\end{figure}
%
%----------------------------------------------------------------------
%----------------------------------------------------------------------
\section{Discussion}
\label{discuss}

\subsection{NGC\,4736 - oval and barred}
\label{discuss1}

The $I$ image (Fig.~\ref{figi}) and the ellipse fits
(Fig.~\ref{figell}) show a clear nuclear bar in M\,94 (NGC\,4736),
with axis ratio $\sim 0.3$ and linear length $\sim 30{\arcsec}$. The
nuclear bar is implicit in the NIR data of Shaw et al (1993) and was
also mentioned by Kormendy (1993). The galaxy NGC\,4736 is known as a
prototypical oval disk galaxy.  Such oval disks can be recognized both
photometrically and kinematically (Kormendy 1982).  Photometrically
the brightness distribution consists of a series of distinct regions,
which each have elliptical isophotes and a noticeable outer edge.  The
axial ratios and position angles of the isophotes in these nested oval
regions are different, implying that at most one of them can be
axisymmetric if they are all coplanar. Kor\-men\-dy (1982) argues that
warps are rare in edge-on spirals at the high surface brightnesses at
which the oval structures occur ($< 25 B$-mag arcsec$^{-2}$) and that
therefore the observed features must be truly non-axisymmetric.
Kinematically, oval disks signify their presence by several features
in their velocity fields which are also known from barred or triaxial
galaxies: The kinematic major axis is different from the photometric
major axis, it is not perpendicular to the kinematic minor axis, and
its position angle may change with radius. These signatures arise from
the projection of quasi-elliptical stellar orbits or gaseous
streamlines orientated at oblique angles with respect to the
line-of-sight and line-of-nodes.

The B-band surface brightness profile of M\,94 was measured by Boroson
(1981); it appears to show several characteristic radii: 40{\arcsec}
and 120{\arcsec}.  Our $I,J,K$ brightness profiles confirm the shelves
at 40{\arcsec} (1.3 kpc) and 120{\arcsec}, although the feature at
40{\arcsec} is less pronounced. The reason for this is probably the
blue color of the star forming ring at $r = 45$\arcsec.  Numerous
large $HII$ regions are located on this inner ring which modify the
blue surface brightness profile (Fig.~\ref{figalph}).  The inner ring
is also seen in $HI$ (at 40-50{\arcsec}, Bosma et al 1977, Mulder \& van
Driel 1994), in CO (Garman \& Young 1986, Gerin, Casoli \& Combes
1991), and in $H_\alpha$ (Fig.~\ref{figalph}, van der Kruit 1974, Buta
1988).  The outer disk of the galaxy is not obviously barred, but the
run with radius of ellipticity and position angle suggests an oval
shape (Fig.~\ref{figell}).  M\,94 is well known for its faint outer
ring at $\sim330${\arcsec} seen in $HI$ and the stellar distribution
(e.g. Plate I in Lindblad 1961 or Fig. 1b in Mulder \& van Driel
1993).  From our kinematic and photometric study we think it is most
likely that the inclination of M\,94 is $i\approx35^{\circ}$, so that
the most axisymmetric region is that portion of the disk between
50{\as} (1.7 kpc) and 100{\as} (3.3 kpc, Sect.~\ref{isoph}).

The kinematics of the $HI$ and $H_\alpha$ gas at medium and large
radii show significant departures from axial symmetry.  A recent
model of the outer oval disk is presented by Gerin et al (1991), who
assume an oval shape of the potential between $5.5\kpc$ and $13\kpc$
and associate the two rings with the inner and outer Lindblad
resonances (ILR and OLR) of this rotating oval. In Sect.~\ref{isoph}
we concluded, that the disk of M\,94 is oval at least between 120{\as}
(4 kpc) and 220{\as} (7.5 kpc); the inner radius is compatible to the
values of Gerin et al, our outer value is lower. However, since the
ellipticity of the disk still increases between 220 and 400{\as}, the
oval disk might well be larger.

The modelling presented in this paper, based on optical CCD
and NIR observations of the central bulge and bar, suggests that the
most likely value for the pattern speed is such that the OLR of the
bar coincides with the position of the inner $HI$ and $H_\alpha$ ring at
45{\arcsec}.  Combining this result with the simulations of Gerin et
al (1991) it thus appears that M\,94 is a galaxy with an inner bar
embedded in an outer oval disk, such that the OLR of the inner
component is located approximately at the ILR of the outer
component. Such bar-within-oval galaxies share many of the
characteristics of the bar-within-bar galaxies discussed, e.g., by
Pfenniger \& Norman (1990) and Friedli \& Martinet (1993). In the
following, we discuss some aspects of this picture for M\,94 in more
detail.

%-----------------------------------------------------------------
\subsection{Nuclear bulge and nuclear bar?}
\label{discuss2}

Does NGC\,4736 have a nuclear bulge?  On the basis of earlier
photometry and kinematic measurements, Kormendy (1993) concluded that
M\,94 is one of a class of galaxies that appear to contain central
bulges which are really only disks. The evidence he presented in
favour of this view was the following: (i) The central component with
the $r^{1/4}$ brightness profile (`bulge') coincides spatially with
spiral structure and a nuclear bar.  These features are characteristic
for disks; thus the $r^{1/4}$ profile cannot {\it per se} be used as
evidence for a truely three-dimensional bulge. (ii) Based on data by
Pellet \& Simien (1982), the apparent $v/\sigma\simeq 0.77$ of
NGC\,4736 is well above that expected for an oblate isotropic rotator
of its flattening ($\varepsilon\simeq 0.1$, $(v/\sigma)_{\rm
obl}\simeq 0.35$). Thus the central kinematics appear disk-like rather
than bulge-like.

With the new data presented in this paper we can now investigate this
question in more detail.  The stellar velocity dispersion profile
(Fig.~\ref{figkin}) shows a clear transition from an outer cold disk
to an inner hot component. Because we are observing M\,94 nearly from
above, the measured dispersion mostly characterizes the vertical
motions in this galaxy. Outside $15{\arcsec}$ the observed dispersion
is consistent with the instrumental resolution ($\sim 60\tkms$) and at
those radii the light may be dominated by a cold disk component. In
this region the circular velocity is $v_c\sim 200\tkms$. Inside
$15{\arcsec}$, the projected velocity dispersion rises to $120\tkms$,
while simultaneously the measured rotation velocity linearly decreases
inwards to zero.

{}From the (admittedly noisy) line-of-sight-velocity distribution ('LOSVD')
shown in Fig.~\ref{figlos}, we see that in fact the rotation velocity
at 10{\as} is still dominated by the narrow peak corresponding to the
rapidly rotating disk. However, the bulge has a much broader, more
symmetric LOSVD with a mean rotation velocity of $\simeq 50\tkms$. This value
reduces {\it the intrinsic bulge} $V_{\rm max}/\sigma$ to a value of
$\sim0.8$ (deprojected), which is more in accord with its {\it
intrinsic} flattening of $q\approx0.6$
(deprojected with $\epsilon=0.1$ and $i=35^\circ$).

If indeed the entire central light came from a constant-thickness
disk, then the observed surface brightness profile, which is
consistent with a projected Hernquist model, would imply a surface
brightness $\propto R^{-1}$, and hence a vertical velocity dispersion
$\propto R^{-1/2}$. Given our limited resolution, the predicted rise
in the vertical dispersion by a factor $\sim2$ between $15{\arcsec}$
and $3{\arcsec}$ is consistent with the measured value of the central
dispersion. However, the shape of the velocity dispersion profile, in
particular the clear decay at around $15{\arcsec}$, argues in favour
of a three-dimensional central bulge component, supporting the
analysis of the LOSVD above.

We next discuss the nature of the central bar.  From the photometry we
infer that the bar comprises only $\ssim 15\%$ of the light within
$20{\arcsec}$. The light distribution is significantly more
axisymmetric in the region $r < 10{\arcsec}$ and $r > 20{\arcsec}$
than near $r = 15{\arcsec}$ where the bar is strongest in the surface
brightness profile (Fig.~\ref{figell}).  The
new surface photometry shows that the bar signature disappears very
quickly for $r > 15{\arcsec}$. The brightness distribution cannot be
fitted with a single ellipsoidal component as the isophotes are far
too pointed near the end of the bar ($a_4$ profile in
Fig.~\ref{figell}). A strong gradient in the ellipticity and boxyness
parameter $a_4$ would be required in a one-component model. This rules
out a single triaxial bulge with nearly constant axis ratio as an
explanation of the data. However, it also argues against a single
disk-like component, for this would have to be almost completely
axisymmetric in the inner $10{\arcsec}$ and develop a separate thin
($b/a\sim 0.3$) bar component near $r = 15{\arcsec}$, a morphology
which is not usually seen in bulge-less, barred galaxies.

Because the projected length of the bar (15\as), the size of the bulge,
and the scale of the linearly rising part of the rotation curve
(Fig.~\ref{figkin}) all coincide, we can not use the velocity dispersion
data to determine the vertical extent of the bar. Indeed, the fact that these
scales approximately agree might suggest that the bar and the bulge
could after all be one and the same component. However, the brightness profile
of the bar is much shallower than that of the bulge, so that in this
interpretation the quadrupole component would have to be strongly
concentrated towards the outer parts of the bulge. Moreover, along
its minor axis direction, the bar is significantly smaller ($\approx 7\as$)
than the bulge (15\as). The former property appears consistent with
the properties of peanut/box-shaped bulges, which -- compared to an
ellipsoidal model -- have more excess flux near their edges than
in their central parts (Shaw 1993). The latter property can not be
compared with these objects because peanut bulges are usually observed
edge-on.

Taken together, the photometry and stellar kinematics of M\,94 therefore
appear to show that this galaxy contains a small central bulge,
a centrally concentrated disk, and a weak $30{\arcsec}$ diameter bar
with shallower profile than either of the former two components.
The bulge may be superposed on the central disk and bar, or, perhaps more
likely, because of the similar surface brightness profiles of
bulge and disk, simply be the inner, more spherical part of the disk.

%-----------------------------------------------------------------
\subsection{Pattern speed of the central bar}
\label{discuss3}

The model discussed in the previous Section assumed that the bar
rotates at a pattern speed $\Omega_p=290 \tkms / \kpc$, such that
corotation lies at $25{\arcsec}$. This places the inner ring of $HI$ and
$H_\alpha$ gas at approximately the OLR of the rotating potential; as
shown by Schwarz (1981) gas between CR and OLR is given angular
momentum by the bar, moves outwards and collects in a ring of material
near the OLR. This zone interior to the OLR is characterized by a weak spiral
pattern, which can best be seen in the morphology of the dust lanes
(Fig.~\ref{figik}). The corotation radius then corresponds approximately
to the end of the bar in the CCD image (Fig.~\ref{figi} and
Fig.~\ref{figbar}), and to the region of lowered cold gas density
visible in the CO observations of Gerin et al (1991).
The depression near corotation is extremely pronounced in the $H_\alpha$
emission distribution (Fig.~\ref{figalph} and Fig.~\ref{fig_gasring}).

Although a model with this value of $\Omega_p$ thus explains a number
of features observed in M\,94, it failed in the sense that the
measured $H_\alpha$ gas kinematics in the central $15{\arcsec}$ are
not fitted by those predicted if the gas moved on closed orbit
streamlines in the barred potential.  However, this may be more a
problem to do with the $H_{\alpha}$ gas: indeed for no reasonable
value of the pattern speed was it possible to obtain a satisfactory
fit to the $H_\alpha$ kinematics.  In any case it is unlikely that the bar
could rotate faster, for then corotation would be placed inside the
bar. There would be no orbits to support the outer bar, yet
the stellar population of the bar appears old ($\gta 3 \gyr$).

Could the bar rotate more slowly? This at least would go in the sense
of somewhat easing the discrepancy of the predicted and the $H_\alpha$
kinematics. In this case the inner ring at $45{\arcsec}$ might be at
ILR rather than at OLR, and then the bar's pattern speed would be
equal to the pattern speed of the outer oval in the Gerin et al
model. If this were the case, the bar would end inside its
ILR. However, such a model could only be in dynamical equilibrium if
the principal axes of bar and oval were either aligned or
perpendicular to each other (Louis \& Gerhard 1988), neither of which
possibility is in accord with the observed misalignment angle between
bar and adjacent disk: From $P.A._{bar}=28^{\circ}$ and
$P.A._{oval}=90^{\circ}$ we get $\Delta\approx62^{\circ}$ in the sky
or $\sim58.5^{\circ}$ in the plane of the galaxy for $i=35^{\circ}$.
If, on the other hand, the two components are oriented at an oblique
angle, then torques between them will have a strong effect
particularly for low pattern speeds. Thus we conclude that the value
of $\Omega_p=290 \tkms / \kpc$ that we have used is the most likely
one.

Models with two non-axisymmetric components rotating at different
pattern speeds were discussed and generated by Pfenniger \& Norman
(1990) and Friedli \& Martinet (1993). These authors concluded that in
order to avoid too large resonant and chaotic zones in the orbit
distribution it was favourable to place the ILR of the outer, more
slowly rotating component onto the CR resonance of the inner, more
rapidly rotating bar. Indeed, in the numerical (N-body and SPH)
simulations of Friedli \& Martinet (1993) two-bar systems with the
corresponding ratio of pattern speeds formed. In the formation of the
inner secondary bar, dissipation is found to be critical in assembling
sufficient amounts of gas onto the $x_2$-orbits in the primary bar
potential.

However, the two-bar phase in these models only lasts for about five
rotations of the faster, inner bar, after which this
dissolves. Therefore, if this is to be a viable scenario for M\,94,
the inner bar must have been triggered very recently, e.g., by
infalling molecular gas, and it must currently end somewhat outside
its ILR.  This appears difficult in the dominant field of a nearly
axisymmetric component.  Moreover, the stars themselves in the M\,94
bar are several times older than the estimated dissolution timescale.
Thus one would require that during their bar triggering no young
stars, formed in the molecular clouds, should have been mixed into the
stellar bar, which appears unlikely.  Alternatively, if the inner bar
indeed rotated faster, such that the ring is at OLR, then the
interaction with the outer oval would be weaker because of the
adiabatic invariants, and the lifetime of this configuration could be
longer.

%-----------------------------------------------------------------
\subsection{Dynamics of the inner $H_\alpha$-emitting gas}
\label{discuss4}

Warm optical emission gas can be detected in M\,94 mainly in
the inner star forming ring ($r \approx 45{\arcsec}$) and in the central
bar region. The dynamics of the ring was studied in detail
by van der Kruit (1974, 1976) and Buta (1988).
The velocity field of the innermost region is fairly complex.
There are several aspects:

(1) The $H_\alpha$ streaming velocities measured in the central
$15{\arcsec}$ of M\,94 are very similar to the stellar streaming
velocities in the bulge (Fig.~\ref{figking}).  These correspond to a
hot stellar system, as the deprojected rotation is $v^2\simeq
\sigma^2/2$ at $10{\arcsec}$ (Sect.~\ref{modkin5}).
Thus the $H_\alpha$ streaming velocities are significantly less than
the circular velocity estimated from the Jeans equation.  However, the
gas rotates faster than the stars for $r > 20{\arcsec}$, fitting to
the HI rotation curves of Mulder
\& van Driel (1993).

(2) The gas velocity field does not reflect the high streaming
velocities expected from closed orbit streamlines in the barred
potential. In particular, the expected steep rise of the measured
velocities in the central $5{\arcsec}$ for slit positions orthogonal
to the bar is not reproduced.

(3) The warm gas velocity field shows, superposed on the general
near-circular rotation field, a number of humps and dips which seem to
correlate with the weak dust spiral pattern observed in the region
between central bar and ring (Fig.~\ref{fighumps}).

We now discuss several possible explanations
for the observed low gas velocities in the very center.

(a) The gas velocities are intrinsically higher, but we do not observe
the high values because of low spatial resolution (seeing and broad slit)
combined with the finite spectral resolution. These effects were taken
into account for the construction of the rotation curves in
Fig.~\ref{fig_rotcurv}, so this possibility can be rejected.

(b) The gas in the region of the central bar moves intrinsically
similar as the stars. This might be plausible if
the observed gas originated from mass loss from the old stars,
and still shared their velocity field.

(c) Walker et al (1988) published $2.0 - 2.5{\mu} m$ spectra from the
central 8{\arcsec} region of M\,94. They showed that the $B_{\gamma}$
emission is very weak there compared to typical starburst galaxies
like M\,82 or NGC\,253. Therefore no strong star formation is presently
going on
in the center, consistent with the not very strong $H_{\alpha}$
emission from the central region. However, from the still relative strong
CO absorption bands of young supergiants Walker et al (1988) concluded
that a strong starburst did occur in the central region of M\,94
some $10^7$ years ago.  A burst of star formation produces wind
bubbles and supernova shocks unrelated to the potential of the galaxy.
Presently, the nuclear region of M\,94 contains a mixture of warm gas,
warm dust, and molecular clouds (Smith et al 1991). The warm gas might
not yet have settled to cloud orbits in the plane, making its present
partly chaotic velocity field plausible.

It will be interesting to compare future high-resolution
molecular gas kinematics with the model streamlines in
Sect.~\ref{modkin}.

%----------------------------------------------------------------------
%----------------------------------------------------------------------
\section{Conclusions}
\label{concl}

(1) We have reported CCD imaging in $I$ and $H_{\alpha}$, focal
reducer imaging in $V,I$, and NIR $J,K$ imaging of the bright Sab
galaxy M\,94 (NGC\,4736).  This galaxy is known as an oval disk
galaxy.  Two rings, an outer ring observed at 330\as and an inner ring
at 45\as, are thought to be located near the outer and inner Lindblad
resonances (OLR and ILR) of this rotating oval (Gerin et al 1991). Our
study of the surface brightness profiles and isophote shapes confirms
the oval disk structure of M\,94 between $r=120\as$ and $r =
220{\arcsec}$. Basic parameters of M\,94 and the most important
numerical results from our work are collected in Table 1.

(2) The central surface brightness distribution of M\,94 is
characterized by a power-law profile. It is not possible to
discriminate bulge and disk components from the photometry alone. We
have constructed a morphological model of the inner regions in terms
of an axisymmetric model with a $\gamma = 0.9$ cusp and scale-length
$\alpha = 15{\arcsec}$ (Dehnen \& Gerhard 1994). Outside $r=60\as$ a nearly
exponential disk with scale-length $\beta \approx 60\as$ must be added
to this model.

(3) Inside the inner ring the position angles of the isophotes
rotate by nearly $90^\circ$. The isophotes show cusped deviations
from ellipses in a range of radii just interior to $r=15\as$.
After subtracting the axisymmetric model, the residuals show a
nuclear bar, with projected radius $r=15{\arcsec}$, axial ratio
$b/a=0.3$ and $P.A.=28^\circ$. The outer oval disk has $P.A.=90^\circ$; the
deprojected misalignment angle between these two components in
the plane of the galaxy is $\Delta = 58.5^{\circ}$.
The luminosity of the inner bar is $5.9 \cdot 10^8 L_\odot$ (in $B$),
comprising about 14\% of the total light inside $20\as$.

(4) Thus M\,94 has a bar-within-bar morphology. We suggest that the
corotation radius of the inner bar is just outside its deprojected
major axis length (20\as = 0.64 \kpc).  This places the inner gas ring
(at 45\as), which is believed to fall near the ILR of the outer oval,
at the OLR of the central bar.  The zone near corotation (CR at 26\as)
is characterized by a marked depression in the $H_\alpha$ and $HI$ gas
densities, consistent with the absence of non-intersecting closed
orbits in this region.  Between CR and OLR of the central bar, spiral
arms are detectable, especially well by their thin dust lanes in,
e.g., $I-K$ color index images.

(5) We have presented absorption line  spectroscopy in 4 longslit
orientations in the inner 60{\arcsec} of M\,94. The stellar velocity
dispersion profile reveals the existence of a central bulge
of $15\as$ radius, with $\sigma=120$\kms. Outside the bulge we observe
$\sigma=60$\kms in the disk region, which is the spectral resolution.
The stellar rotation curves are smooth and are consistent with an
axially symmetric velocity field with inclination $i=30^\circ$ and
a line of nodes at $P.A.=113^\circ$. This suggests that the effects of the
nuclear bar on the stellar velocity field are weak, in the bulge region
because this is a hot stellar system, and outside the bar because there the
potential becomes rapidly axisymmetric.  The combination of stellar
rotation and velocity dispersion is consistent with the photometric
model in the bulge and adjacent disk regions in the sense that the
Jeans equation is satisfied for $M/L_B \approx 1.8$.

(6) Futher evidence that the bulge and disk components in M\,94 are
distinct comes from the line-of-sight velocity profiles at positions
along the minor axis but beyond the extent of the nuclear bar, which
suggest a rapidly rotating, cold component superposed onto a more slowly
rotating, hot component. From these profiles we estimate that the
deprojected mean rotation of the bulge is $v\approx100$\kms, and that
$v/\sigma=0.8$, roughly consistent with an intrinsic bulge axis ratio
$c/a=0.6$.

(7) By contrast, the kinematic data do not allow us to determine
whether the bulge and bar components are distinct.  The result that
the radius of the bulge and the major axis length of the bar are
similar might suggest that, in fact, bulge and bar are one and the
same component, perhaps similar to that inferred in the inner Galaxy.
However, the bar is significantly smaller along its minor axis
direction than the bulge, so that a simple near-ellipsoidal component
is not a good model for the data.

(8) We have obtained emission line spectroscopy in 8 longslit orientations
of the inner $100\as$ of M\,94, and thus obtained new results on the
gas kinematics in the region of the inner bar. The streaming velocity
field of the $10^4\rm K$ gas is similar to that of the stars for $r<20\as$.
The gas rotates faster than the stars and fits the HI rotation field
(Mulder and van Driel 1993) at larger radii, consistent with circular orbits.

(9) Orbit calculations in the gravitational potential derived from the
photometry of M\,94 predict large non-circular motions for cold gas in
equilibrium in the bulge region. The resulting kinematics do not
reproduce the observed central $H_\alpha$ kinematics. For the future
it would be extremely interesting to compare high-resolution
CO gas velocity observations with the model
predictions.

(10) The result that the $H_\alpha$ streaming velocities in the inner
$15\as$ are less than required for gravitational equilibrium suggests
that the warm ionized gas has significant random motions. This could
be because this gas is still associated with the stars through recent
mass loss, or, perhaps more likely, that it is beeing heated by shocks
and localized gas flows. Indeed, while Walker et al (1988) found no
$B_{\gamma}$ in the central 8{\arcsec} in M\,94, i.e.
no signs  for an on-going starburst, they concluded from the strengths
of the CO bands that a strong starburst must have occurred several
$10^7$ yr ago. Direct evidence for some still ongoing star formation
in the inner regions of M\,94 also comes from the observed
$H_{\alpha}$ emission.

%----------------------------------------------------------------------
\acknowledgements
%________________________________________ Do not leave a blank line here!
We thank the Calar Alto observatory staff for the support during the
observations. We thank the MAGIC crew (Heidelberg) for the efficient
support during the NIR campaign.  We thank C. Scorca (Heidelberg) and
P. Surma (Cambridge) for the exposure of the focal reducer images and
F. Baier and H. Tiersch (Potsdam) for the exposure of the $H_{\alpha}$
image. We are grateful to R. Kohring (Heidelberg) for substantial
support during the reduction of the emission line spectrograms.  We
acknowlegde the use of the Fourier-cross-correlation routine of
R. Bender (M\"unchen) and the $\gamma$-routine of W. Dehnen
(Oxford). The SIMBAD data base in Strasbourg was used to search for
references of M\,94. This work was supported by the {\em Deut\-sche
For\-schungs\-ge\-mein\-schaft} via the {\em
Sonder\-for\-schungs\-be\-reich 328}.

%----------------------------------------------------------------------
%----------------------------------------------------------------------

%%%%%%%%%%%%%%%%%%%%%%%%%%%%%%%%%%%%%%%%%%%%%%%%%%%%%%%%%%%%%%%%%%%%%

\begin{table*}
\caption[ ]{Parameters of M\,94}
\begin{flushleft}
\begin{tabular}{l l l l}
\hline
&Item			&Quantity		       &Reference\\
\hline\\
%-------------------------------------------------------------------------
{\bf General}
&type			&(R)Sab(s) resp. (R)SA(r)ab II &RSA resp. RC2\\
{\bf Parameters}
&distance		&6.6 Mpc		       &\\
&scale			&$1{\as} = 32pc$	       &\\
&$v_{sys}$		&310 km/s		       &Buta 1988\\
&mass			&$5.1\times10^{10}M_{\sun}$
                                                &Mulder {\&} van Driel 1993\\
&$B_t$			&8.96 mag		       &Buta 1988\\
&			&			       &\\
%-------------------------------------------------------------------------
{\bf Morphology}
&bulge			&$r = 15-20$\as		       &Sect. 3.7\\
{\bf on Sky}
&central bar		&$a=15\as$, $b=7\as$, $P.A.=28^{\circ}$ &Sect. 3.1\\
&inner ring		&$r = 45$\as, $P.A.=127^{\circ}, \varepsilon=0.27$
                                                &van der Kruit 1976\\
&oval disk		&$r =120-220\as$, $P.A.\approx95^{\circ},
                        \varepsilon=0.23$             &Sect. 3.2\\
&outer ring		&$r \approx 330$\as, $P.A. \approx 150^{\circ}$
                                                      &Buta 1988\\
&			&			      &\\
%-------------------------------------------------------------------------
{\bf Inner Disk+Bulge}
&brightness profile     &$\gamma=0.9, \alpha=15$\as   &Sect. 3.2\\
{\bf Surface Photometry}
&ellipticity		&$\varepsilon \approx 0.18$   &Sect. 3.4\\
&position angle		&$P.A. \approx 110^{\circ}$   &Sect. 3.4\\
&inclination		&$i \approx 35^{\circ}$       &Sect. 3.4\\
&mass of inner disk	&$7.0\times10^{9}M_{\sun}$    &Sect. 5.1\\
&			&			      &\\
%-------------------------------------------------------------------------
{\bf Bulge Model}
&flattening		&$q=0.4$			      &Sect. 3.7\\
&mass			&$2.0\times10^{9}M_{\sun}$    &Sect. 5.1\\
&$v/\sigma$		&0.8			      &Sect. 3.7\\
&$M/L_B$		&1.8			      &Sect. 5.1\\
&			&			      &\\
%-------------------------------------------------------------------------
{\bf Central Bar Model}
&axis lengths (deprojected)	&$a=21\as$, $b=7\as$	      &Sect. 5.1\\
&mass 			&$9.8\times 10^8M_{\sun}$     &Sect. 5.1\\
&pattern speed		&290 km/s/kpc		      &Sect. 5.4\\
&ILR inner Lindblad res.&$5{\as}\cor160pc$	      &Sect. 5.4\\
&CR  corotation radius  &$26{\as}\cor830pc$	      &Sect. 5.4\\
&OLR outer Lindblad res.&$40{\as}\cor1280pc$	      &Sect. 5.4\\
&			&			      &\\
%-------------------------------------------------------------------------
{\bf Stellar Kinematics}
&$v_{rot}$ (deprojected)&230km/sec max. at 15\as      &Sect. 5.2\\
&velocity dispersion	&120 km/sec		      &Sect. 3.7\\
&inclination		&$i=30^{\circ}$		      &Sect. 3.7\\
&line of nodes		&$P.A.=113^{\circ}$	      &Sect. 3.7\\
&			&			      &\\
%-------------------------------------------------------------------------
{\bf Gas Kinematics}
&$v_{rot}$ (deprojected) &250km/sec max. at 15-20\as  &Sect. 3.8\\
&			&			      &\\
%-------------------------------------------------------------------------
{\bf $HI$}
&mass           	&$10^9M_{\sun}$	         &Mulder {\&} van Driel, 1993\\
&$v_{rot}$		&$130-170km/s$	         &Mulder {\&} van Driel, 1993\\
&inclination		&$i\approx40^{\circ}$	 &Mulder {\&} van Driel, 1993\\
&line of nodes		&$P.A.\approx115^{\circ}$&Mulder {\&} van Driel, 1993\\
&			&			 &\\
%-------------------------------------------------------------------------
{\bf Molecular Gas}
&$H_2$ mass		&$5.1\times10^8M_{\sun}$ &Gerin et al 1991\\
&$v_{rot}$		&200 km/s at 60\as 	 &Gerin et al 1991\\
&			&			 &\\
\hline
\end{tabular}
\end{flushleft}
\end{table*}
%-------------------------------------------------------------------------

\end{document}